\documentclass[
aip,
 amsmath,amssymb,
 reprint,%
floatfix]{revtex4-1}

\usepackage{graphicx}
\usepackage{dcolumn}
\usepackage{bm}

\usepackage[utf8]{inputenc}
\usepackage[T1]{fontenc}
\usepackage{mathptmx}
\usepackage{etoolbox}
\usepackage{svg}
\usepackage{textcomp}
\usepackage{soul}  
\usepackage[mathlines]{lineno}
\makeatletter
\def\@email#1#2{%
 \endgroup
 \patchcmd{\titleblock@produce}
  {\frontmatter@RRAPformat}
  {\frontmatter@RRAPformat{\produce@RRAP{*#1\href{mailto:#2}{#2}}}\frontmatter@RRAPformat}
  {}{}
}%

\def \paren#1{\left( #1 \right)}
\def \brack#1{\left[ #1 \right]}
\def \angle#1{{\langle #1\rangle}}

\makeatother
\begin{document}
\preprint{AIP/123-QED}

\title[\textit{Accepted for Publication in the Review of Scientific Instruments (DOI: 10.1063/5.0326026): For Personal Use Only}]{The EXoplanet Climate Infrared TElescope (EXCITE):\\A balloon-borne mission to measure spectroscopic phase curves of transiting hot Jupiters}
\author{T.~D.~Rehm}
 \email{timothy.d.rehm@nasa.gov}
 \affiliation{NASA Goddard Space Flight Center, Greenbelt MD, USA}
\author{C.~Altermatt}
 \affiliation{Department of Physics, Brown University, Providence RI, USA}
\author{L.~Bernard}
 \affiliation{School of Earth and Space Exploration, Arizona State University, Tempe AZ, USA}
\author{A.~Bocchieri}
 \affiliation{Department of Physics, La Sapienza Universit\`a di Roma, Rome, Italy}
\author{N.~Butler}
 \affiliation{School of Earth and Space Exploration, Arizona State University, Tempe AZ, USA}
\author{O.~Carey}
 \affiliation{Department of Physics, Brown University, Providence RI, USA}
\author{R.~C.~Challener}
 \affiliation{Department of Astronomy, Cornell University, Ithaca NY, USA}
\author{J.~Hartley}
 \affiliation{StarSpec Technologies Inc., Cambridge Ontario, Canada}
\author{K.~R.~Helson}
 \affiliation{NASA Goddard Space Flight Center, Greenbelt MD, USA}
 \affiliation{Center for Space Sciences and Technology, University of Maryland, Baltimore County, Baltimore MD, USA}
\author{D.~P.~Kelly}
 \affiliation{NASA Goddard Space Flight Center, Greenbelt MD, USA}
\author{K.~Klangboonkrong}
 \affiliation{Department of Physics, Brown University, Providence RI, USA}
\author{A.~L.~Korotkov}
 \affiliation{Department of Physics, Brown University, Providence RI, USA}
\author{M.~Lally}
 \affiliation{Department of Astronomy, Cornell University, Ithaca NY, USA}
\author{E.~Leong}
 \affiliation{NASA Goddard Space Flight Center, Greenbelt MD, USA}
\author{N.~K.~Lewis}
 \affiliation{Department of Astronomy, Cornell University, Ithaca NY, USA}
\author{S.~Li}
 \affiliation{StarSpec Technologies Inc., Cambridge Ontario, Canada}
\author{M.~Line}
 \affiliation{School of Earth and Space Exploration, Arizona State University, Tempe AZ, USA}
\author{S.~F.~Maher}
 \affiliation{NASA Goddard Space Flight Center, Greenbelt MD, USA}
 \affiliation{Triggs Tech, LLC, Ellicott City MD, USA}
\author{R.~McClelland}
 \affiliation{NASA Goddard Space Flight Center, Greenbelt MD, USA}
\author{L.~V.~Mugnai}
 \affiliation{School of Physics and Astronomy, Cardiff University, Cardiff, UK}
\author{P.~C.~Nagler}
\affiliation{NASA Goddard Space Flight Center, Greenbelt MD, USA}
\author{C.~B.~Netterfield}
 \affiliation{StarSpec Technologies Inc., Cambridge Ontario, Canada}
 \affiliation{Department of Physics, University of Toronto, Toronto Ontario, Canada}
\author{V.~Parmentier}
 \affiliation{Department of Physics, University of Oxford, Oxford, UK}
\author{E.~Pascale}
 \affiliation{Department of Physics, La Sapienza Universit\`a di Roma, Rome, Italy}
\author{J.~Patience}
 \affiliation{School of Earth and Space Exploration, Arizona State University, Tempe AZ, USA}
\author{L.~J.~Romualdez}
 \affiliation{StarSpec Technologies Inc., Cambridge Ontario, Canada}
\author{P.~A.~Scowen}
 \affiliation{NASA Goddard Space Flight Center, Greenbelt MD, USA}
\author{G.~S.~Tucker}
 \affiliation{Department of Physics, Brown University, Providence RI, USA}
\author{I.~Waldmann}
 \affiliation{Department of Physics and Astronomy, University College London, London, UK}

\date{\today}

\begin{abstract}
The EXoplanet Climate Infrared TElescope (EXCITE) is a balloon-borne mission dedicated to measuring spectroscopic phase curves of hot Jupiter-type exoplanets. Phase curve measurements can be used to characterize an exoplanet's longitude-dependent atmospheric composition and energy circulation patterns. EXCITE carries a 0.5 m primary mirror and moderate resolution diffraction-limited spectrograph with spectral coverage from 0.8--3.5\,\textmu m. EXCITE is designed to fly from a long-duration balloon (LDB). EXCITE will observe through the peak of a target's spectral energy distribution (SED) and through spectral signatures of hydrogen and carbon-containing molecules. In this paper, we present the science goals of EXCITE, detail the as-built instrument, and discuss its performance during a 2024 engineering flight from Fort Sumner, New Mexico.
\end{abstract}

\maketitle


\section{\label{sec:intro}Introduction}

This paper presents an overview of the EXoplanet Climate Infrared TElescope (EXCITE).~\cite{nagler2022exoplanet, pascale2021exoplanet, tucker2018exoplanet} EXCITE is a balloon-borne instrument designed to measure the spectra of transiting extrasolar giant planets, also known as ``hot Jupiters,'' in the near-infrared (NIR). EXCITE carries a 0.5\,m diameter primary mirror and a moderate resolution NIR spectrograph with spectral coverage from 0.8--3.5\,\textmu m. Hot Jupiters orbit close to their host stars, less than 0.1\,AU, and have orbital periods of $\sim$\,1 to 5~days. EXCITE will continuously measure spectra of these planets throughout their entire orbits, an observational technique known as phase-resolved spectroscopy.\cite{stevenson2014thermal} EXCITE is designed to operate from a long-duration balloon (LDB) platform deployed near Earth's poles and will measure phase-resolved spectra of known hot Jupiters. The duration of an LDB flight is on average >\,1~week; the longest on record was 58~days from the Southern Hemisphere. In a single flight, EXCITE can significantly increase the number of measured spectroscopic phase curves of hot Jupiters. EXCITE flew a test flight on August 31, 2024 from the NASA Columbia Scientific Balloon Facility (CSBF) in Fort Sumner, New Mexico, and the mission is currently preparing for an Antarctic LDB flight. CSBF is the NASA balloon launch facility and operates out of Palestine, Texas. CSBF also has balloon launch locations in New Mexico, Oregon, Sweden, New Zealand, Australia, and Antarctica.  This paper is organized as follows. Section~\ref{sec:science} gives an overview of EXCITE's science goals and the motivation for flying from an LDB platform. Section~\ref{sec:architecture} introduces the EXCITE architecture and details each as-built subsystem. Section~\ref{sec:Simulations} gives an overview of EXCITE's simulation and data reduction tools and presents a study of instrumental noise and systematic effects. Section~\ref{sec:FTS} discusses the results of EXCITE's 2023 and 2024 flight campaigns. Our conclusions are in Section~\ref{sec:summary}. 

\section{\label{sec:science}EXCITE Science}

\begin{figure}
    \centering
    \includegraphics[width=\linewidth]{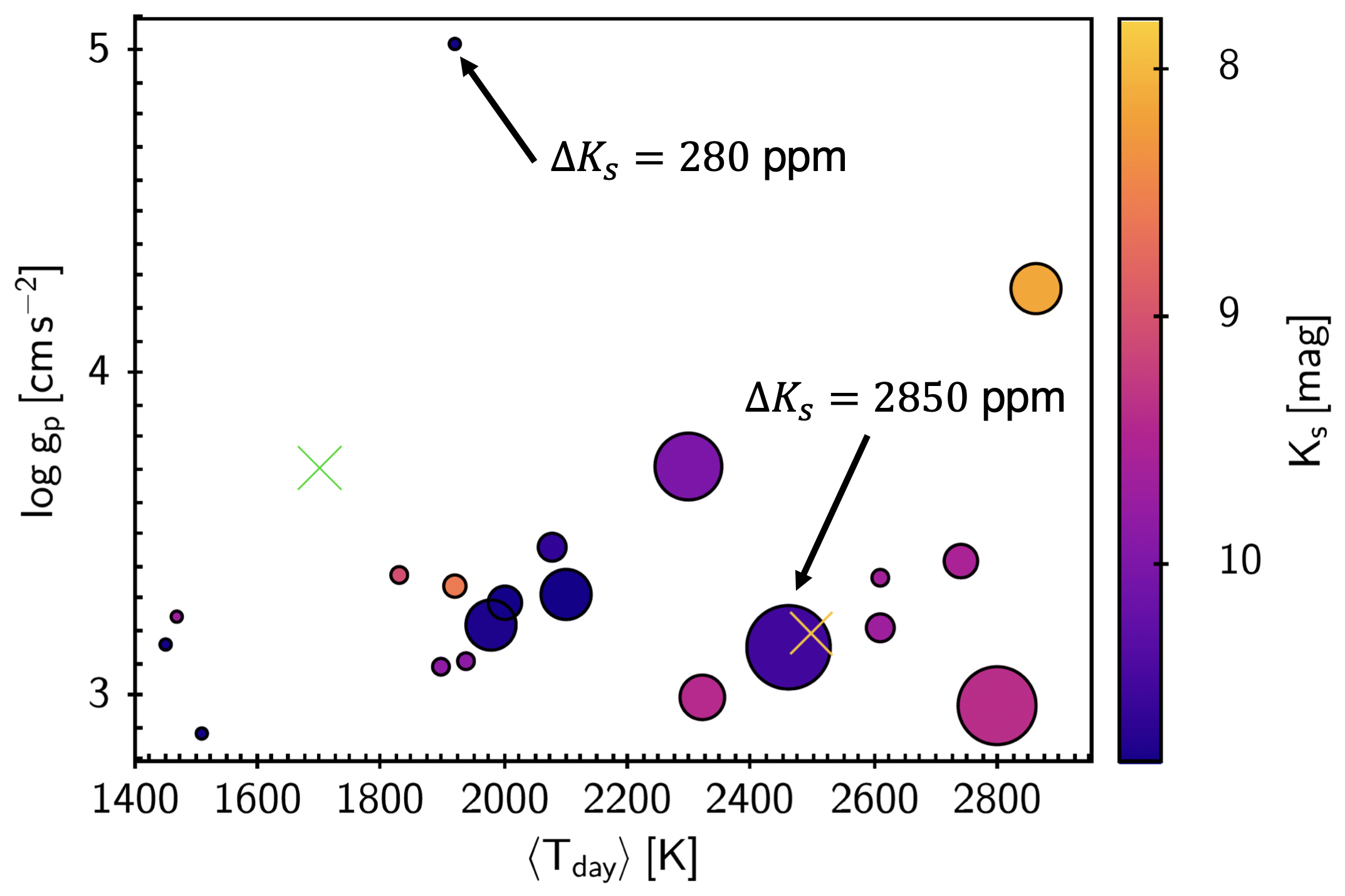}
    \caption{Surface gravities ($\log$\,g$_{\text{p}}$) and dayside temperatures $\angle{T_{\text{day}}}$ of potential EXCITE targets and their K$_{s}$-band magnitudes.~\cite{nagler2022exoplanet} These targets are available to observe during a single Antarctic LDB flight. Measuring a diverse set of targets in this parameter space allows for spectral classification of hot Jupiters.~\cite{nagler2019observing} Targets with lower surface gravity are good candidates for transit spectroscopy since the limb of the atmosphere is larger. Targets with higher dayside temperature tend to have large secondary eclipse depths since the thermal emission of the exoplanet is large. The X's indicate WASP-43b~\cite{stevenson2014thermal} and WASP-103b,~\cite{kreidberg2018global} the first two exoplanets for which phase-resolved spectroscopy has been published. Reproduced with permission from Proc. of SPIE Vol. 12184, 121840V (2022). Copyright 2022 SPIE.}
    \label{fig:paramspace}
\end{figure}

EXCITE aims to perform phase-resolved spectroscopy of transiting hot Jupiters. Hot Jupiters are a class of exoplanets that are good candidates for atmospheric studies. Since they have orbital periods of only a few days, it is possible to measure many full-orbit phase curves of different targets in a relatively short period of time. These exoplanets are tidally locked with their host stars, resulting in large temperature gradients between their daysides and nightsides. This gives rise to strong equatorial jets, transporting heat and material around the planet.~\cite{guillot1996giant} Hot Jupiters are strongly irradiated to temperatures between 1000--4000\,K, resulting in spectral energy distributions (SEDs) that peak between 0.7--3\,\textmu m. EXCITE will measure phase curves of hot Jupiters throughout entire orbits and through the peak of their SEDs. Phase curve measurements determine the change in the brightness of an exoplanet-star system as a function of orbital phase. These observations naturally include two notable events: the transit and the secondary eclipse. The transit occurs when an exoplanet passes in front of its host star. For hot Jupiters, the exoplanet typically blocks about 1\%, or $10^{4}$ parts-per-million (ppm), of the light from the host star during transit. This results in a dip in the total brightness measured from the system. The amplitude of the decrease in brightness is called the ``transit depth." The secondary eclipse occurs when an exoplanet passes behind its host star, resulting in another dip in the measured brightness. The amplitude of this dip is called the ``secondary eclipse depth.'' This amplitude is smaller than the transit depth and occurs because the host star blocks both the thermal emission from the exoplanet and reflected starlight. In the NIR, the thermal emission component typically dominates. In this case, the secondary eclipse depth, $D_{e}$, can be calculated as follows:
\begin{equation}\label{eq:depths}
    D_{e}(\lambda) = \paren{\frac{R_{p}}{R_{\star}}}^{2}\frac{BB\paren{\angle{T_{\text{day}}}, \lambda}}{BB\paren{T_{\star}, \lambda}},
\end{equation}
where $\angle{T_{\text{day}}}$ is the mean dayside temperature of the exoplanet, $T_{\star}$ is the temperature of the star, $R_{p}$ is the radius of the exoplanet, $R_{\star}$ is the radius of the host star, and $BB$ refers to the blackbody radiation at a given temperature and wavelength. $\angle{T_{\text{day}}}$ is given to good approximation by~\cite{Seager+2010}
\begin{equation}\label{eq:tday}
    \angle{T_{\text{day}}} = T_{\star}\sqrt{\frac{R_{\star}}{a}}\brack{f'\paren{1-A_{B}}}^{1/4}.
\end{equation}
Here, $a$ is the semi-major axis, $A_{B}$ is the Bond albedo, and $f'$ is a variable which parametrizes the heat recirculation efficiency. For hot Jupiters with poor heat recirculation efficiency, $f'=2/3$ and $A_{B}\sim 1/4$.~\cite{Seager+2010, schwartz2017phase} In the NIR, the secondary eclipse depth for hot Jupiters ranges between 200--3000\,ppm. Figure~\ref{fig:paramspace} shows a sample parameter space for targets observable by EXCITE during an Antarctic LDB flight. Targets with secondary eclipse depths in the K$_{s}$-band ($\sim$\,2.2\,\textmu m) greater than 250\,ppm are shown. To achieve sufficient signal-to-noise in observing the dimmest targets in our list ($D_{e}\sim$\,280\,ppm), we target instrument noise below 40\,ppm. We show that we achieve this sensitivity in Section~\ref{sec:Simulations}. By probing targets that sample the parameter space shown in Figure~\ref{fig:paramspace}, EXCITE aims to create a spectral classification of hot Jupiters based on their temperature, size, composition, and brightness.

Photometric transit and secondary eclipse measurements can provide accurate measurements of a hot Jupiter's orbital period, radius, and effective temperature.~\cite{parmentier2017exoplanet} Spectroscopic measurements of a hot Jupiter during transit and secondary eclipse can be used to characterize its atmosphere. During transit, when the cooler nightside faces the line of sight, stellar light is filtered, scattered, and absorbed by the exoplanet atmosphere, resulting in an absorption spectrum when measured. The mass of an exoplanet can be constrained by measuring the transmission spectrum, since the pressure profile, and resulting absorption profile, is heavily dependent on the exoplanet's mass.~\cite{de2013constraining} At phases close to the secondary eclipse, the hotter side of the exoplanet is visible. The planet's emission produces excess flux on top of the stellar spectrum. Emission spectra can reveal the vertical pressure-temperature and chemical abundance profiles of the hottest part of an exoplanet's atmosphere. EXCITE will measure absorption spectra during transit and emission spectra through secondary eclipse of multiple hot Jupiters during one LDB flight.

There are limitations to observing only the transit and/or the secondary eclipse. Spectroscopic transit observations are best suited for targets with low surface gravities since these planets tend to have the larger limbs. Transit spectra of targets with high surface gravities more difficult to measure. Similarly, spectra and brightness maps produced from secondary eclipse observations are only hemispherical averages from one side of the exoplanet. In order to characterize the global atmospheric properties of a planet, a full-orbit phase curve connecting the two events is required. When observing phase curves, the observer measures flux from both the star and the exoplanet; the phase curve can be parameterized by the planet-to-star flux ratio, $F_{p}/F_{\star}$, where $F_{p}$ is the flux from the exoplanet and $F_{\star}$ is the flux from the star. The amplitude of the planet-to-star flux ratio defines the sensitivity required to distinguish between flux levels at different phases between transit and secondary eclipse. We model $F_{p}/F_{\star}$ as a sinusoidal variation with a phase curve amplitude of $(F_{d}-F_{n})/2F_{\star}$, where $F_{d}$ is the dayside flux of the planet, and $F_{n}$ is the nightside flux of the planet. $F_{p}/F_{\star}$ is written as
\begin{equation}\label{eq:Fp/Fstar}
    \frac{F_{p}}{F_{\star}}(t,\lambda) = \frac{F_{d}+F_{n}}{2F_{\star}}+\frac{F_{d}-F_{n}}{2F_{\star}}\cos\paren{\frac{2\pi t}{P_{p}}+\phi_{o}}\sin(i),
\end{equation}
where $P_{p}$ is the orbital period, $\phi_{o}$ is the phase offset, and $i$ is the inclination angle of the exoplanet. Equation~\ref{eq:Fp/Fstar} does not include the transit and secondary eclipse. In the case of tidally locked hot Jupiters with poor heat recirculation from the dayside to the nightside, we assume that $F_{d}\gg F_{n}$, and the phase curve amplitude and constant offset equal $D_{e}(\lambda)/2$. The phase offset is a measure of the longitudinal difference between the substellar point and the planet's hottest hemisphere. EXCITE does not spatially resolve the star and planet, and therefore produces a single spectrum for each target system. Since $F_{\star}\gg F_{p}$, EXCITE only measures small perturbations in the stellar spectrum, but we plan to continuously measure through two secondary eclipses per observation, which allows us to measure baseline stellar spectra without contributions from the planet.

Photometric phase curves provide measurements of the day- and nightside temperatures through the phase curve amplitude and the phase offset. Together, these quantities constrain the exoplanet's advective timescale (which depends on wind speed and drag), radiative timescale, Bond albedo, and heat redistribution efficiency.~\cite{schwartz2017phase} The phase offset is also sensitive to the presence of clouds in the atmosphere,~\cite{parmentier2021cloudy} particularly if there are reflective clouds on the dayside.~\cite{coulombe2025ltt9779} Phase offsets in hot Jupiters are not uncommon and have been observed in WASP-14b,~\cite{wong20153} WASP-19b,~\cite{wong20163} HAT-P-7b,~\cite{wong20163} HD 149026b,~\cite{knutson20098} CoRoT-2b,~\cite{dang2018detection} and more. The majority of these hot Jupiters display eastward offsets due to fast equatorial jets. From this subset, only CoRoT-2b exhibits a westward shift of the brightest spot, where $F_{p}/F_{\star}$ peaks after the secondary eclipse. Because hot Jupiters are tidally locked, there is a unique hemisphere and longitude of the exoplanet facing the observer at each phase. Thus, phase curve observations can be inverted into longitudinally-resolved brightness and temperature maps of the atmosphere's photosphere.~\cite{stevenson2014thermal}

Spectroscopic phase curves probe the wavelength dependence of the planet-to-star flux ratio as a function of longitude. As with photometric observations, these phase curves can be used to create brightness maps that reveal atmospheric dynamics. However, the wavelength dependence of these maps contains a wealth of additional information. 
Spectroscopic phase curves can be used to create longitudinal maps at different altitudes, as different wavelengths probe different pressures, dependent on the atmospheric composition and longitudinal variations in the vertical temperature structure.~\cite{stevenson2014thermal} In general, wavelengths that probe the upper atmosphere (\textit{i.e.}, wavelengths with significant absorption) will show smaller phase offsets, particularly if there is significant atmospheric drag (\textit{e.g.}, magnetic drag~\cite{beltz2022magnetic}), while the deeper atmosphere will show larger phase offsets. EXCITE's broad wavelength coverage provides access to a large range of atmospheric opacity sources and the associated changes in the depth of the photosphere, thereby constraining atmospheric dynamics throughout the atmosphere. By performing phase-resolved spectroscopy, we can break degeneracies between atmospheric signatures and planetary thermal emission, while also constraining the global energy budget.

Longitudinally-resolved emission spectra reveal the location-dependent chemistry of the exoplanet atmosphere, which is expected to vary across the planet due to changes in temperature, irradiation, and atmospheric dynamics. EXCITE's spectral range spans the peak of the hot Jupiter SED, allowing us to constrain the temperature of the exoplanet. The NIR spectral passband of EXCITE extends through absorption features of many molecules expected to be, or already observed, in hot Jupiters. H$_2$O, commonly detected in hot Jupiters,~\cite{coulombe2023w18b, yang2024w43b, gressier2025w17b} has numerous signatures across the EXCITE passband. For ultra-hot Jupiters with dayside temperatures >\,2500\,K, the H$_2$O abundance is reduced through thermal dissociation,~\cite{arcangeli2018h} and H$^-$ is produced through hydrogen dissociation.~\cite{bell2018h2diss} At shorter wavelengths (1--1.5\,\textmu m), H$^{-}$ becomes a large opacity source and reduces the spectral features of H$_{2}$O, resulting in an emission profile which closely resembles a blackbody spectrum. This limits the information derivable from emission spectra between 1--2\,\textmu m. The interpretation of phase curves at these wavelengths can be ambiguous since the night side of an exoplanet can appear dark either because the total flux is low or because there is strong absorption. By observing through the peak of the SED, EXCITE will resolve this degeneracy and unambiguously measure the planet’s global energy budget. Beyond 2\,\textmu m, features from molecules that do not dissociate, such as CO, are visible, as is the resurgence of H$_{2}$O lines.~\cite{parmentier2018thermal, lothringer2018extremely} The spectral reach of the EXCITE passband will allow it to improve upon phase curve data measured by \textit{Hubble Space Telescope} (\textit{HST}) Wide Field Camera 3 (WFC3), which is limited to 1.1--1.7\,\textmu m. Possible TiO and VO abundances have been measured in WASP-121b.~\cite{evans2017ultrahot} These molecules typically exist at high altitudes and are large ultraviolet absorbers; their presence can lead to thermal inversions in an exoplanet's atmosphere.~\cite{gandhi2019new} While evidence for inversions exist,~\cite{hubeny2003possible,fortney2008unified} an instrument with EXCITE's sensitivity and bandpass is required for confirmation.

Phase curve measurements are vital for providing empirical constraints to General Circulation Models (GCMs). GCMs compute fully-3D atmospheres and are a powerful tool to interpret observations. In general, GCMs provide a good match to secondary eclipse data,~\cite{showman2009atmospheric} but more recent work shows they can struggle to replicate the shapes of phase curve observations.~\cite{kempton2023gj1214b, bell2024w43b} The GCMs used to fit phase curve data would greatly benefit from observations that bridge the 1.7--3.5\,\textmu m gap between spectroscopic data taken by \textit{HST}/WFC3 and photometric data taken by \textit{Spitzer Space Telescope}. EXCITE will greatly expand the number of observed phase curves that can be used to refine GCMs to better understand exoplanet atmospheric dynamics.

\begin{figure}
\centering
\includesvg[width=\linewidth]{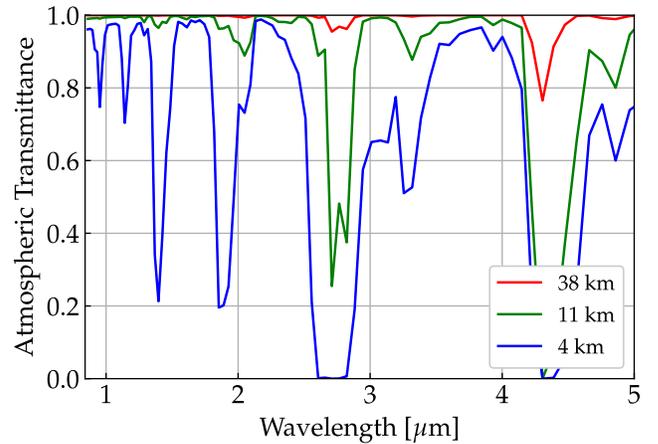}
\caption{\label{fig:transmittance}
Atmospheric transmittance as a function of wavelength for observations from an LDB platform (red), a commercial aircraft (green), or at the height of Mauna Kea on the ground (blue). The transmittance profiles are binned to a spectral resolution of 50.}
\end{figure}

Spectroscopic phase curve observations, while well motivated scientifically, are resource intensive and challenging in practice. An observing platform must have a broad spectral coverage in order to measure molecular features of many atmospheric constituents, as well as be located in space or near-space to avoid significant telluric contamination. Currently, the only instrument located in space with broad enough spectral coverage to measure through the peak of a hot Jupiter's SED is the \textit{James Webb Space Telescope} (\textit{JWST}). However, the low spectral-resolution observing mode (PRISM) of the Near-Infrared Spectrograph (NIRSpec) is the only \textit{JWST} instrument that simultaneously covers the complete EXCITE passband, and this observing mode is J-mag limited to targets dimmer than 9.83,~\cite{nielsen2016jwst} ruling out the best candidates for atmospheric studies. EXCITE will be able to measure spectra of the brightest hot Jupiters across 0.8--3.5\,\textmu m at once. Although more limited in its spectral coverage, \textit{HST}/WFC3 is located in low Earth orbit (LEO) and can also measure full-orbit spectroscopic phase curves. These measurements are still challenging because the observing environment is dynamic, and the relative position between the instrument and the Sun varies quickly, leading to rapid changes in the ambient temperature between -100\,$^{\circ}$C and 100\,$^{\circ}$C. This results in long thermal settling times, upwards of the duration of an entire \textit{HST} orbit after a slew.~\cite{debes2019pushing} Similarly, phase curves measured with \textit{HST}/WFC3 introduce large systematic ``observation gaps'' that necessitate careful detrending and multiple visits during one observation. Therefore, measuring phase curves from LEO typically requires observation times much longer than the orbital period of the exoplanet. For example, \citet{arcangeli2019climate} assembled a full-orbit phase curve of WASP-18b, which has a 22.5-hour orbital period, using 18.5 orbits of \textit{HST}. \citet{stevenson2014thermal} needed $\sim$\,60 \textit{HST} orbits to measure three complete orbits of WASP-43b, which has a 19.5-hour orbital period. \citet{kreidberg2018global} required 46 hours of \text{HST} observing time and an additional 30 hours of \textit{Spitzer} observing time to perform phase resolved spectroscopy across 1.1--1.7\,\textmu m range and photometry at 3.6 and 4.5\,\textmu m. At a practical level, the long-duration stares required for phase curve measurements are seen as burdensome for the limited, and shared, resources of space-based instruments. Because of this, only a handful of full-orbit spectroscopic phase curves have been observed: WASP-43b,~\cite{stevenson2014thermal, Bell2023JWST, challener2024latitudinal} WASP-103b,~\cite{kreidberg2018global} WASP-18b,~\cite{arcangeli2019climate} and WASP-121b.~\cite{mikal2023jwst, splinter2025precise} These hot Jupiters were all observed with shared space-based instruments: \textit{HST}, \textit{JWST}, and \textit{Spitzer}. Significantly increasing the number of measured spectroscopic phase curves is better suited for a dedicated instrument.

The broad spectral coverage required for NIR phase curve measurements is not available from the ground or aircraft, but is accessible to stratospheric balloons. Figure~\ref{fig:transmittance} shows Earth's atmospheric transmittance across 0.8--5\,\textmu m for several observing platforms at different altitudes. EXCITE will fly from an LDB platform at stratospheric altitudes between 38--40\,km, above 99.5\% of the Earth's atmosphere, and measure across 0.8--3.5\,\textmu m at a spectral resolution of at least $R\sim 50$. At these altitudes and spectral resolution, the atmosphere is stable and nearly transparent. In Section~\ref{sec:ELDiP}, we discuss how we subtract the background contribution from the Earth's atmosphere from spectral images. Because the target exoplanets are always observable from Earth's poles, it will be possible to obtain phase-resolved spectra through continuous ``point-and-stare'' observations, with the photometric stability required to measure the phase curve amplitudes. Because the environment is also thermally stable, systematic observation gaps and their associated settling times are eliminated. In one LDB flight, the EXCITE mission will substantially increase the number of full-orbit spectroscopic phase curves of hot Jupiters.

\section{\label{sec:architecture}The EXCITE Payload}

\begin{figure}
\centering
\includegraphics[width=\linewidth]{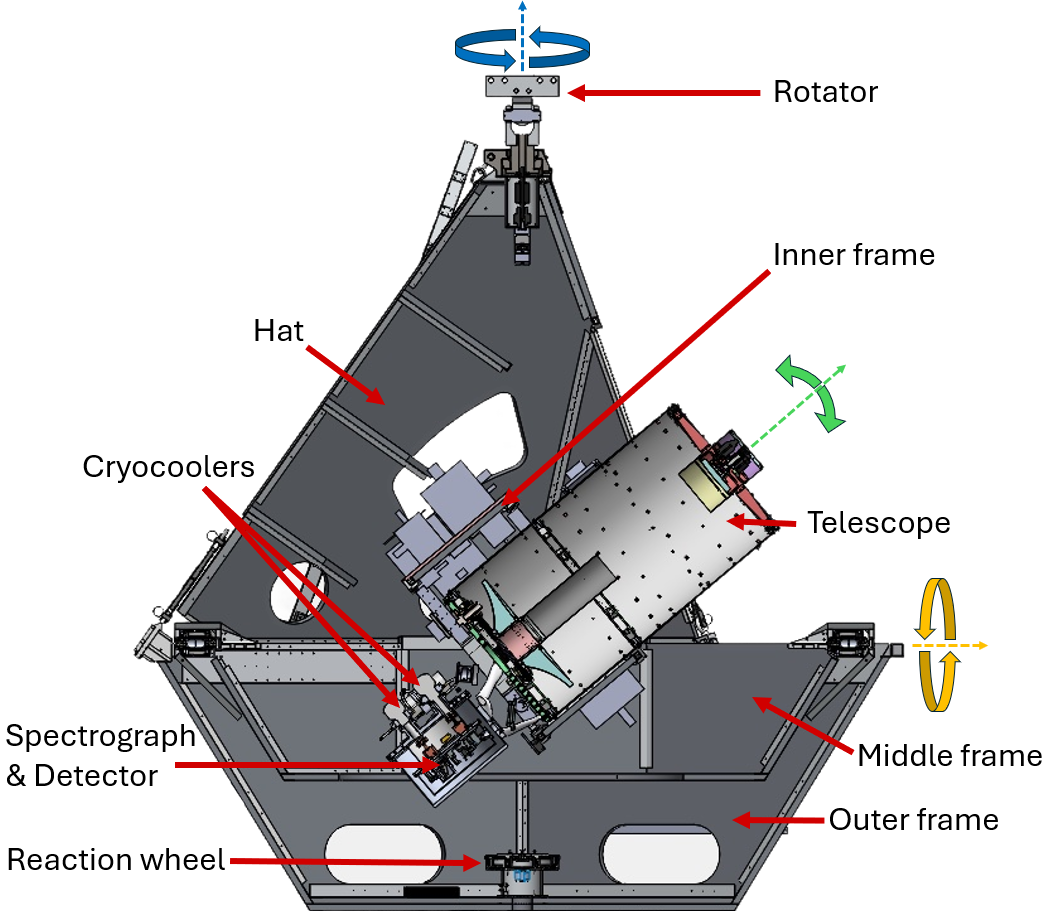}
\caption{\label{fig:EXCITEgondola} A diagram of the EXCITE gondola and science instrument. The science instrument is located in the middle of the gondola. The nested frames and other components are labeled. The outer frame connects to the hat/rotator and rotates in the azimuth direction shown with the blue arrows. Roll rotations occur at the middle frame and rotate in the direction shown by the orange arrows. The inner frame actuates the science instrument in the elevation direction shown with the green arrows.}
\end{figure}

\begin{figure}
    \centering
    \includegraphics[width=\linewidth]{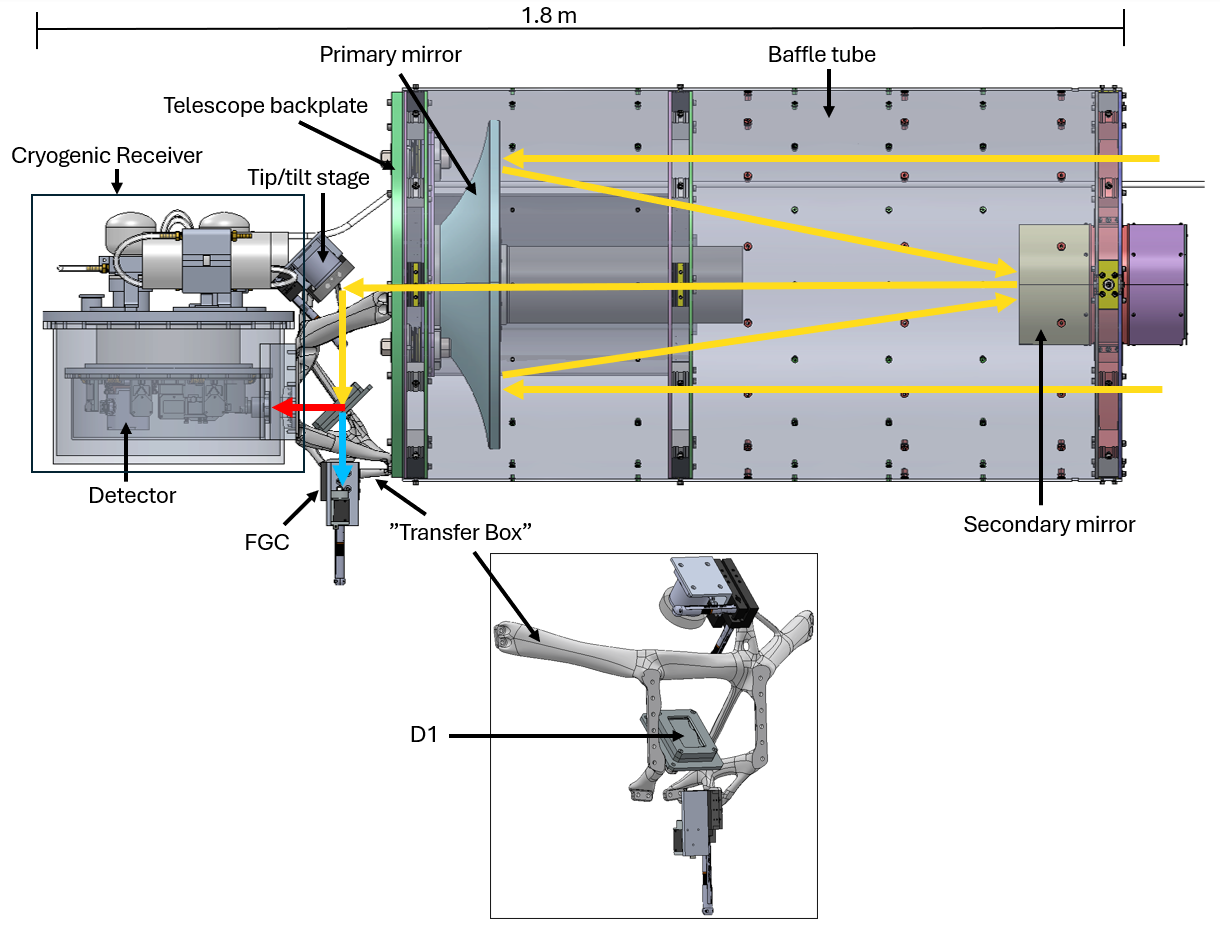}
    \caption{A CAD model of the EXCITE science instrument showing the optical path. Light collected by the telescope (yellow arrows) reflects off the primary and secondary mirrors to a piezo-actuated tip/tilt mirror. This mirror reflects incident light to an ambient temperature dichroic. Here, the light is split into a 0.6--0.8\,\textmu m transmitted component (blue arrow) and a 0.8--3.5\,\textmu m reflected component (red arrow). The transmitted beam is collected by an optical camera (FGC) which provides pointing feedback to the tip/tilt stage, enabling high line-of-sight stability ($\sim$\,50~mas rms) at the telescope focus. The reflected component is the science beam. It passes through the entrance window of the cryostat and into the spectrograph. The spectrograph collimates the beam, disperses the light with a prism, focuses the dispersed beam, and splits it into short (0.8--2.5\,\textmu m) and long (2.5--3.5\,\textmu m) wave channels that are then incident on the detector. A zoomed-in region below the instrument shows the ``transfer box.'' The transfer box opto-mechanically couples the cryogenic receiver to the telescope, and acts as the optical bench for the fine guidance system (FGS). Its design is described by McClelland.~\cite{mcclelland2022generative} A baffle (not shown) is incorporated into the transfer box and limits optical stray light on the FGC.}
    \label{fig:instrument}
\end{figure}

The EXCITE payload is composed of a balloon gondola and a science instrument (Figures~\ref{fig:EXCITEgondola} and~\ref{fig:instrument}). The gondola consists of a mechanical structure that holds the science instrument and an attitude control system (ACS). The science instrument is made up of a telescope, fine guidance system (FGS), and a cryogenic receiver. The cryogenic receiver comprises a cryostat, a slit-less NIR spectrograph, and a HgCdTe detector system with associated readout electronics. The gondola, ACS, and telescope are near-copies of those used for the Super-pressure Balloon-borne Imaging Telescope (SuperBIT) mission.~\cite{gill2024superbit} The SuperBIT gondola has flown a total of five combined engineering and LDB flights and has demonstrated gondola stabilization of <\,1\,$''$ rms and line-of-sight stabilization of 50~milliarcseconds (mas) rms at the focal plane.~\cite{romualdez2020robust}

Here we detail EXCITE's as-built subsystems. Section~\ref{sec:gondola} describes the gondola and ACS. Section~\ref{sec:telescope} describes the telescope and FGS. Section~\ref{sec:cryo} describes the cryogenic system. Section~\ref{sec:spectrograph} describes the spectrograph. Section~\ref{sec:detector} describes the detector system.

\subsection{\label{sec:gondola}Gondola \& ACS}

The EXCITE gondola consists of three gimballed frames that can each be stabilized to better than 1\,$''$ (1-$\sigma$). The gondola is fabricated from aluminum honeycomb, providing high mechanical stiffness in a lightweight configuration. Romualdez~\cite{romualdez2018design} presents a full description of the SuperBIT gondola, ACS, and the onboard software which manages commanding, telemetry, core computations, and power distribution control. EXCITE's analogous systems are based upon this architecture.

\begin{figure}
    \centering
    \includegraphics[width=\linewidth]{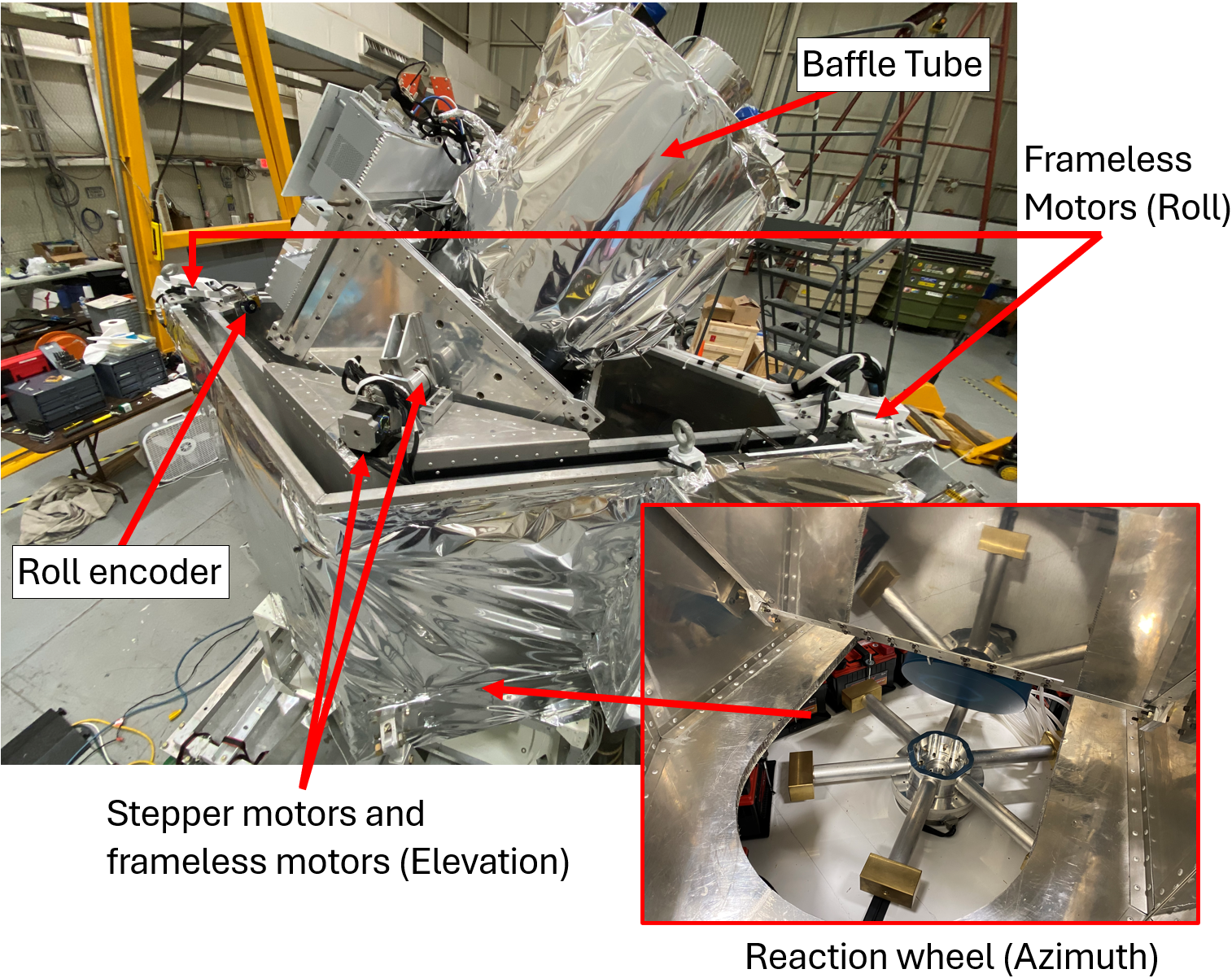}
    \caption{The EXCITE payload with its hat removed. The starboard side of the payload is displayed; the stern direction is towards the left, and the bow direction is towards the right. The port side of the payload is obscured. The reaction wheel (bottom) stabilizes in azimuth. The roll motors are located at the bow and stern, where the middle frame is attached to the outer frame. The elevation actuators use both coarse stepper motors and frameless motors. They are located on the port and starboard sides where the telescope and inner frame attach to the middle frame. The roll encoder is located on the stern frameless motor, and the elevation encoder is located on the port side and cannot be seen. The gyroscopes mount to the telescope baffle tube and are covered by aluminized Mylar.}
    \label{fig:gimbalmotors}
\end{figure}

Figure~\ref{fig:EXCITEgondola} shows a diagram of the EXCITE gondola. The ACS stabilizes the science instrument's line-of-sight in three nearly-orthogonal axes: azimuth, elevation, and roll. Azimuthal motion is achieved using a high-inertia reaction wheel located inside the outer frame and a rotator mechanism at the top of the gondola to dump angular momentum to the balloon flight train. EXCITE can point in any azimuthal direction. We plan to point $\pm$60$^{\circ}$ anti-Sun during science operations in flight. Elevation motion is achieved using two pairs of stepper and frameless motors that work in tandem. The stepper motors are responsible for coarse motion ($1.8^{\circ}$ resolution), and the frameless motors for fine motion, down to the sub-arcsecond scale. Accessible elevation angles are limited to 22$^{\circ}$--57$^{\circ}$ to avoid the ground and the balloon; mechanical hard stops enforce this constraint. The frameless motors have $\pm$6$^{\circ}$ travel. When tracking on a target, the stepper motors keep the frameless motors near the center of their travel. Adjusting the roll angle (rotations about the bow-stern axis) corrects for field rotations during observations. Roll actuation is provided by a pair frameless motors similar to those used for fine elevation motion. Their travel is also limited to $\pm$6$^{\circ}$. Figure~\ref{fig:gimbalmotors} displays the mechanisms which are used to stabilize the gondola. When the roll frameless motors reach the limits of their travel, they are reset in a process that lasts $\sim$\,30\,s.
 
\begin{figure}
    \centering
    \includegraphics[width=\linewidth]{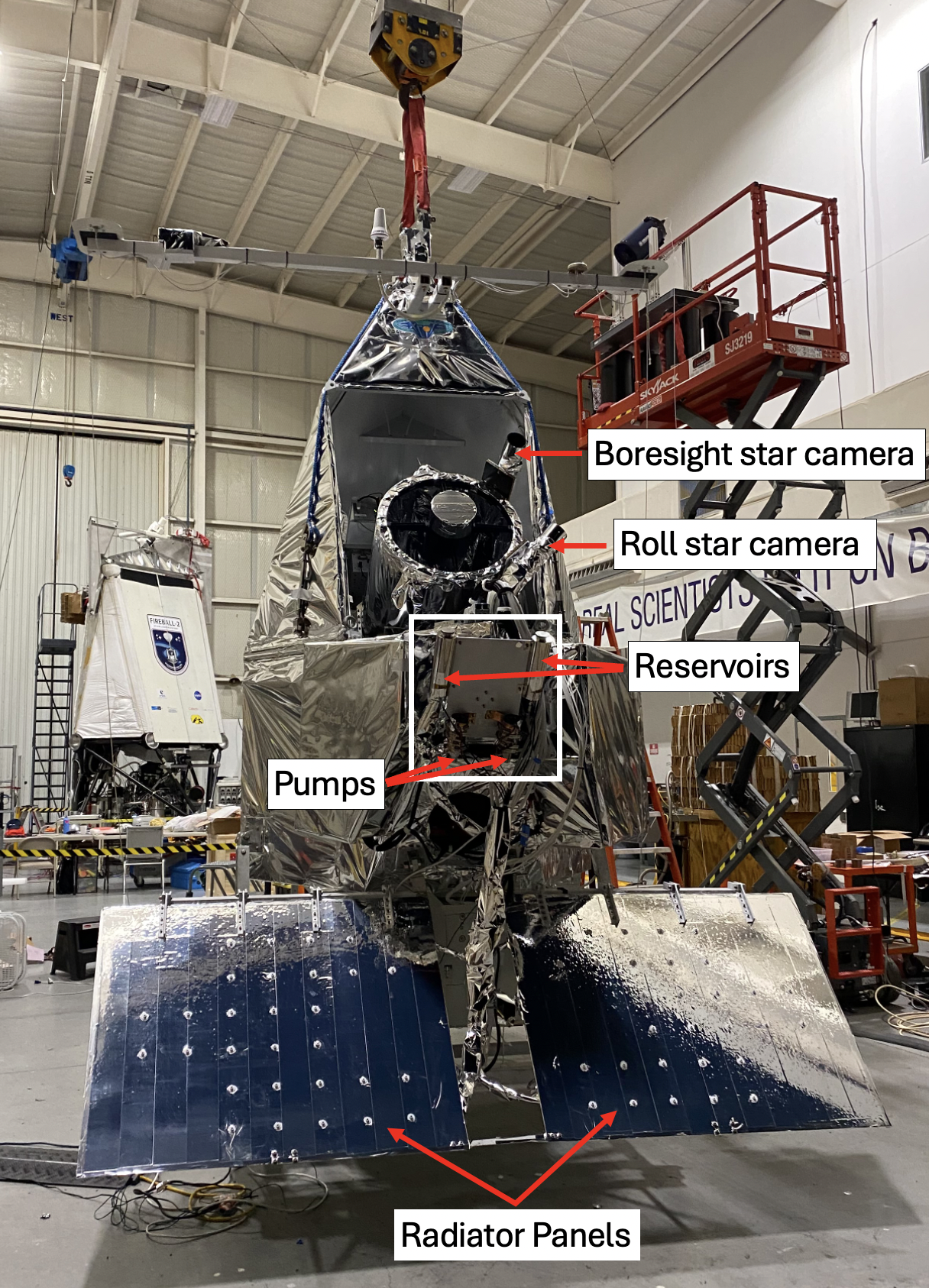}
    \caption{The EXCITE payload during an on-sky pointing test in Fort Sumner, New Mexico in 2023. The locations of the star cameras are labeled. Also shown are the locations of the fluid pumps, methanol reservoirs, and radiator panels used by the cryogenic system (see Section~\ref{sec:cryo}.)}
    \label{fig:EXCITE_FTS}
\end{figure}

Feedback to the mechanisms is provided by a suite of sensors: three fiber optic rate gyroscopes, two star cameras, a magnetometer, and two rotary encoders. The encoders report the angular position of the elevation and roll axes. The gyroscopes and magnetometer supply inertial and positional feedback to the reaction wheel and rotator mechanism and dampen low frequency disturbances along the azimuthal axis. The gyroscopes read out angular velocities about three orthogonal axes. Two gyroscopes are located on the telescope baffle tube, and one is located on the backplate of the telescope. Once the gondola is stabilized with feedback from the gyroscopes and encoders, two orthogonal star cameras determine absolute positioning. The star cameras are located on the telescope baffle tube and point along the boresight and about the roll axis. Figure~\ref{fig:EXCITE_FTS} shows the location of the star cameras on the telescope. Each camera has a 4$^{\circ}$\,$\times$\,4$^{\circ}$ field of view and provides absolute pointing information via multi-star imaging and ``lost-in-space'' pattern recognizing algorithms. The cameras are capable of high-resolution centroiding at the sub-arcsecond level ($\sim$\,0.23\,$''$ and $\sim$\,0.46\,$''$ centroids on boresight and roll, respectively) at a 0.5\,Hz rate.~\cite{romualdez2018overview} The star cameras also act as a jitter feedback and correction mechanism when they are locked-on to bright stars. Altogether, the gondola and ACS provide three-axis target acquisition and stabilization down to <\,1\,$''$.

\begin{figure}
    \centering
    \includegraphics[width=\linewidth]{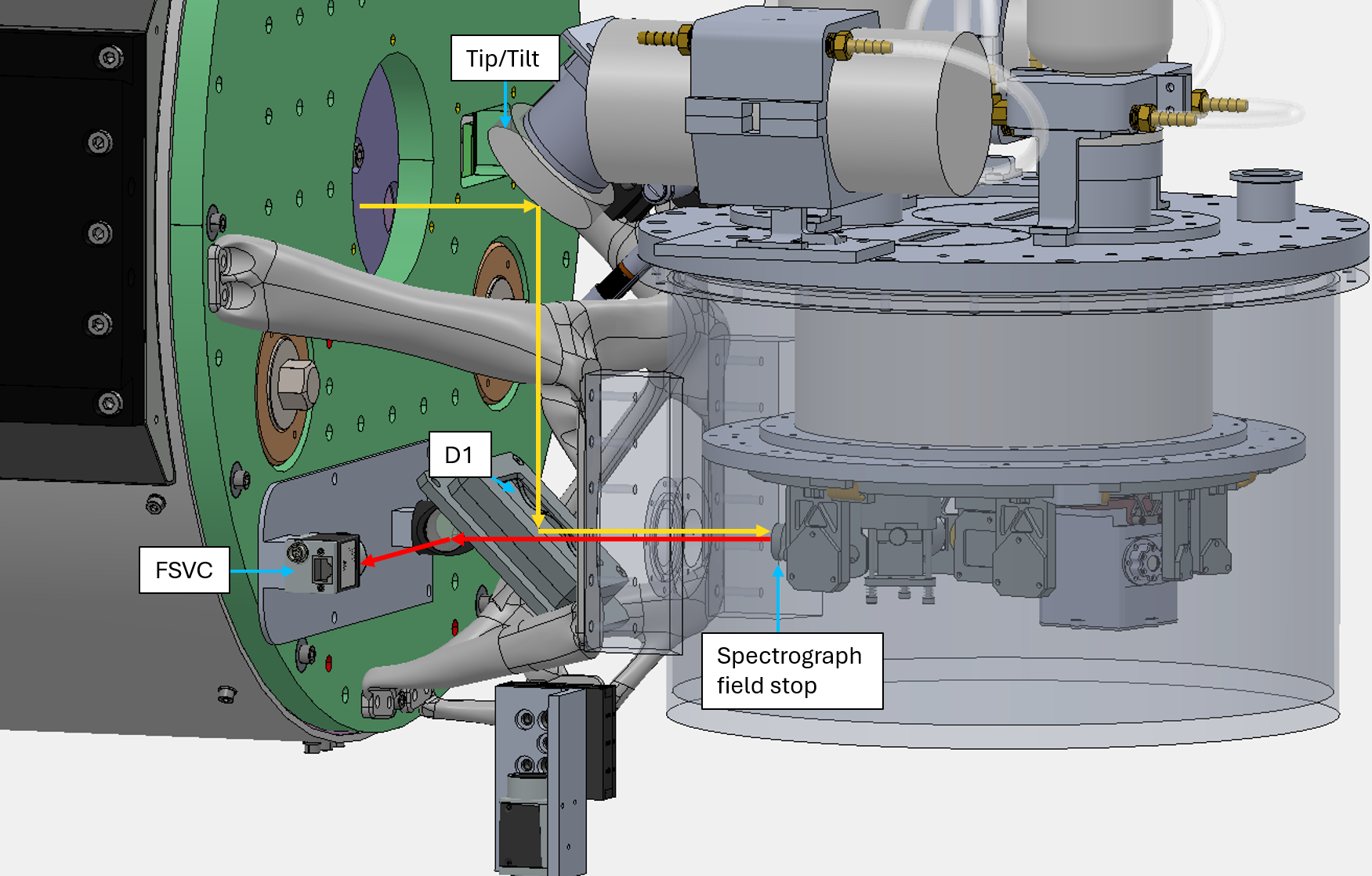}
    \caption{An outline of the optical configuration of the FSVC. Light focused by the telescope (yellow arrows) emerges through the telescope backplate (shown in green) and reflects off the tip/tilt mirror. The light is guided to D1, and the reflected portion enters the receiver through  the field stop. Light that does not pass through the field stop will reflect off its substrate (red arrows); some portion will travel back through D1. A fold mirror guides this lights into the FSVC. Both the fold mirror and the camera are mounted to the backplate of the telescope. The blue arrows are used to accompany the labels of different components and do not represent any relevant light path.}
    \label{fig:fsvc}
\end{figure}

\subsection{\label{sec:telescope}Telescope \& FGS}

The EXCITE telescope has a 0.5\,m diameter primary mirror and a 0.13\,m diameter secondary mirror in an $f/12$ Ritchey-Chr\'etien configuration. The primary and secondary mirrors are made from fused silica and have a protected silver coating. While protected gold can offer higher reflectivity in the NIR, we chose protected silver due to limitations by the telescope vendor. The telescope has an on-axis wavefront error <\,60\,nm rms, surface roughness <\,10\,nm rms, and average optical throughput >\,95\% across 0.8--4\,\textmu m. A 0.64\,m diameter carbon fiber baffle tube surrounds the telescope and supports the secondary mirror. The inside of the baffle tube is lined with five concentric aluminized Mylar blankets. The Mylar blankets reduce the NIR thermal emission from the ambient temperature baffle tube and enable passive cooling of the telescope optics. Three actuators are located behind the secondary mirror to adjust its tip, tilt, and focus. These can be used on the ground and in flight. Two dovetail plates fastened to opposite sides of the baffle tube are used to fix the telescope to the gondola's inner frame. During assembly, these plates allow the instrument $\pm0.15$\,m of translational freedom along the boresight axis to allow for precise center-of-mass balancing. The outside of the telescope baffle tube contains an array of threaded inserts that are used to mount the roll and boresight star cameras, the gyroscopes, balancing weights, and cables. To limit throughput of off-axis stray light, an inner baffle extends from the primary mirror aperture, shown in Figure~\ref{fig:instrument}. 

The FGS consists of a piezo-actuated tip/tilt mirror and an optical camera. The transfer box (Figure~\ref{fig:instrument}) acts as the FGS's optical bench. Light that is collected by the telescope reflects off the tip/tilt mirror and illuminates an ambient temperature dichroic (D1). The transmitted component is incident on an optical camera at the telescope focus, known as the ``fine guidance camera'' (FGC). The FGC provides feedback to the tip/tilt mirror. The FGS operates independently from the ACS and stabilizes the line-of-sight at the telescope focus down to $\sim$\,50~mas at a bandwidth of $\sim$\,1--40\,Hz. The science focal plane can also supply feedback to the tip/tilt stage to reduce longer timescale drifts. The FGC is mounted to a translation stage that allows focusing on the ground and in flight.

A Field Stop-Viewing Camera (FSVC) helps aim the science beam into the spectrograph, accounting for any differential misalignment between the tip/tilt stage and D1. This camera is mounted on the back of the telescope and images the field stop through the substrate of D1, monitoring the location of the focused science beam on the field stop substrate. We include a small wedge angle in D1 to compensate for lateral shifts to the science beam. The FSVC is not part of the telescope optical path and it is not in the pointing control loop. Rather, it is used to calibrate the spectrograph entrance location on the FGC. The FSVC is equipped with a zoom lens and a bandpass filter centered at 800\,nm. A light-emitting diode (LED) is mounted adjacent to the zoom lens and can be used to illuminate the field stop substrate to aid in focusing the FSVC. A figure of the optical path and geometry used by the FSVC is shown in Figure~\ref{fig:fsvc}.

\subsection{\label{sec:cryo}Cryogenic System}

\begin{figure}
    \centering
    \includegraphics[width=\linewidth]{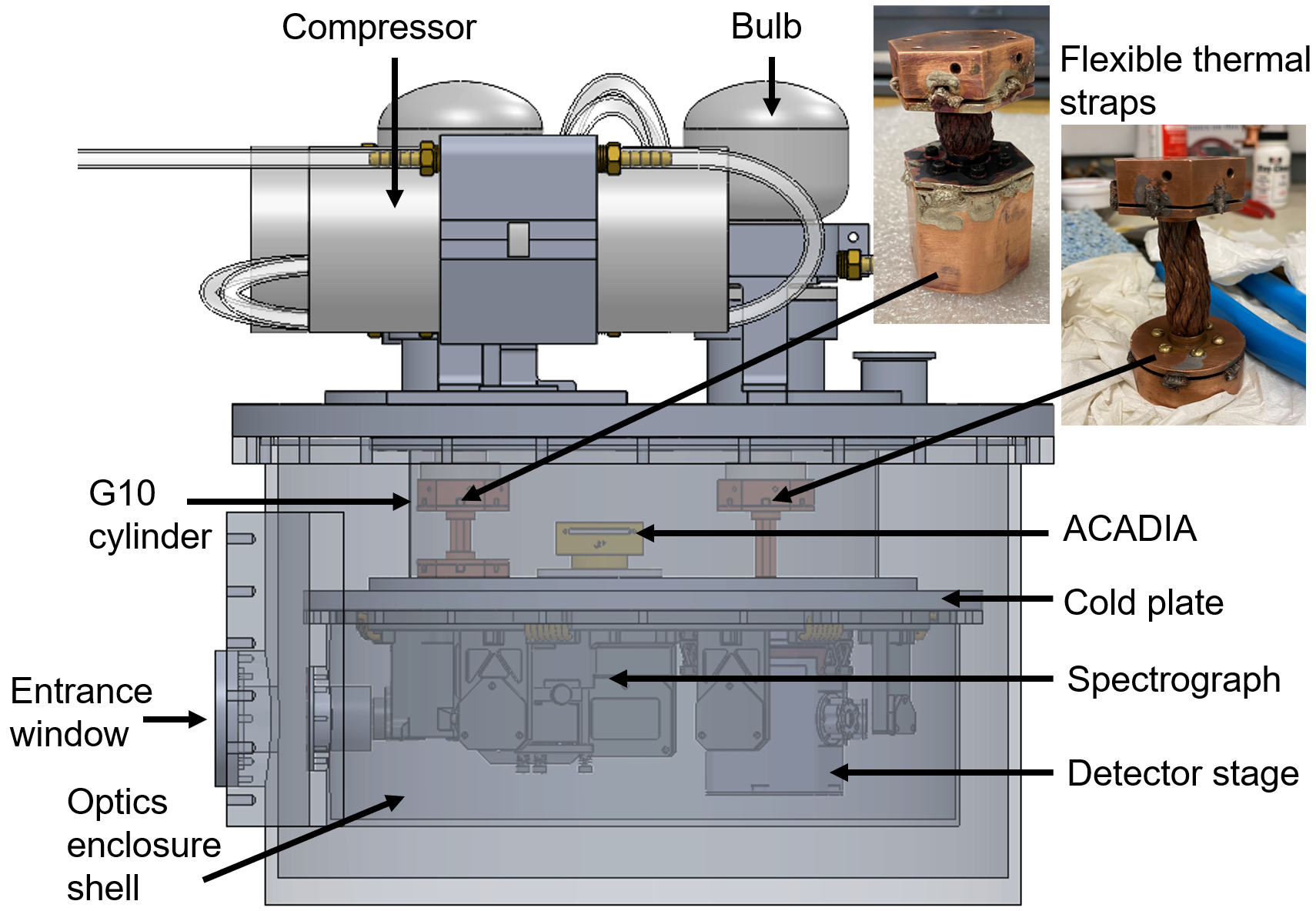}
    \caption{\label{fig:cryostat_CAD} A transparent side view of the internal structure of the EXCITE cryostat. The top right shows photos of the custom thermal straps made to the attach the cryocoolers to their respective stages.}
\end{figure}

\begin{figure}
    \centering
    \includegraphics[width=\linewidth]{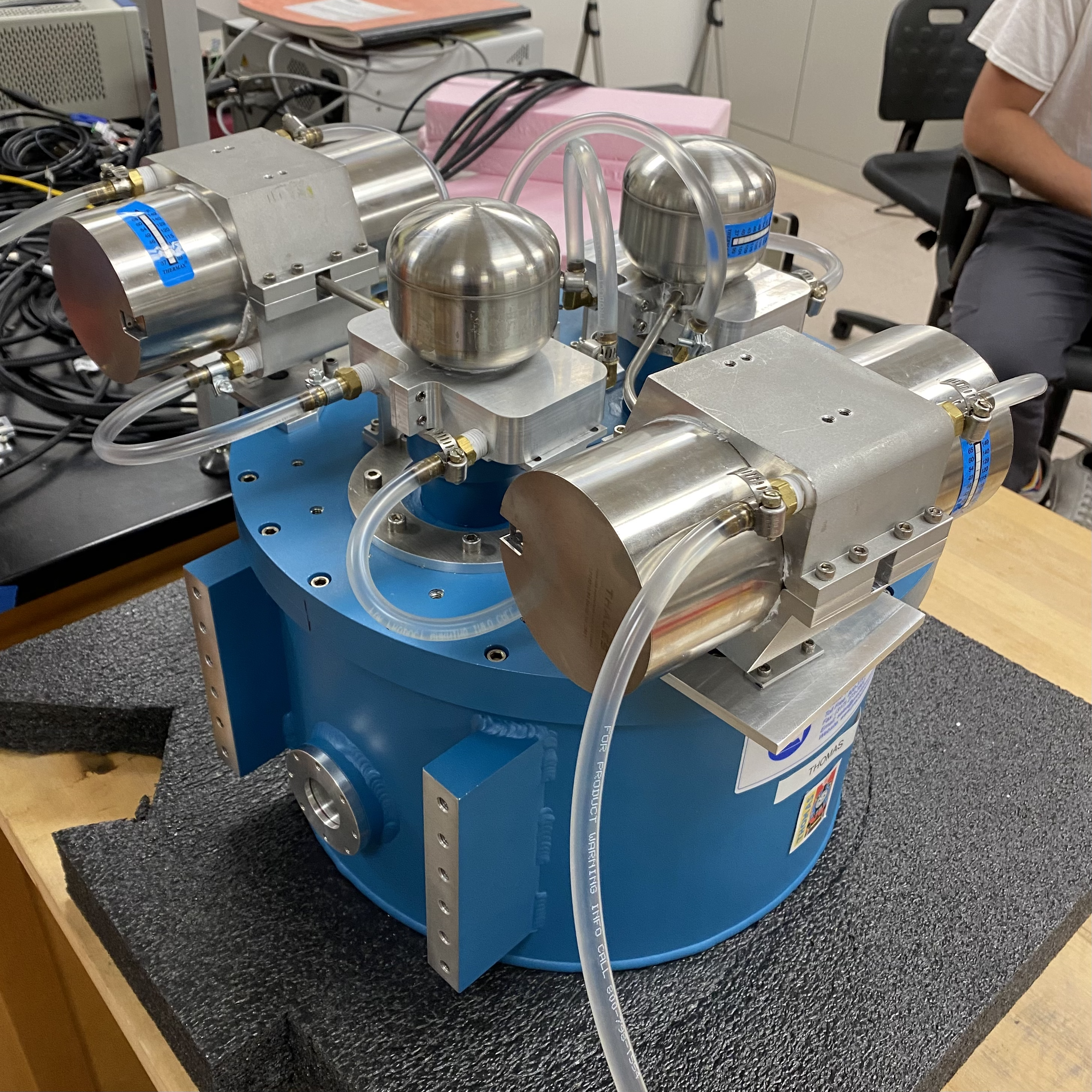}
    \caption{\label{fig:cryostat_real} The EXCITE cryostat assembled with two Thales LPT9310 cryocoolers installed in their heat sinks. Each cryocooler has a bulb and a compressor. Power dissipated by the coolers is split roughly evenly between the bulbs and compressors. The 100\,K cryocooler comprises the leftward compressor and the bulb closest to the front of the image. The 50\,K cryocooler comprises the rightward compressor and the bulb in the back of the image.}
\end{figure}

The EXCITE cryogenic system is composed of a cryostat with a dewar and cryocoolers, control electronics, and a fluid loop that dissipates heat generated by the cryocoolers.~\cite{rehm2022design, rehmthesis2025} The cryostat is $\sim$\,35.6\,cm diameter $\times$ 30.5\,cm tall cylinder that mounts to the transfer box. It contains the spectrograph, detector, and ASIC for Control And Digitization of Imagers for Astronomy (ACADIA) readout electronics. Figure~\ref{fig:cryostat_CAD} shows the internal elements of the EXCITE cryostat, and Figure~\ref{fig:cryostat_real} shows a photo of the cryostat with the cryocoolers and heat sinks. The entrance aperture is a 31.75\,mm diameter cutout, and the entrance window is made of 3\,mm-thick anti-reflection coated sapphire with >\,90\% transmission across the 0.8--3.5\,\textmu m band. Internally, an aluminum cold plate is kept at 120\,K and is mechanically fixed to the top plate of the cryostat through a G10 thermally-isolating cylinder. The optical bench of the spectrograph is fastened to the underside of the cold plate. The ACADIA mounts to the top side of the cold plate and is thermally controlled to maintain a stable temperature near 130\,K. The detector is mounted to a copper cold plate that is fixed to the spectrograph's optical bench via a series of low thermal conductivity Kevlar suspension structures. Its temperature is also actively controlled. A 50 K box with a pair of apertures covers the detector. The spectrograph and detector box are enclosed by an optics shell which is also cooled to 120\,K. The inner surface of the optics shell is painted with black Aeroglaze Z306 to minimize stray light at the focal plane and to passively cool the optical elements. Its outer surface is wrapped in multi-layer insulation (MLI) to reduce radiative loading on the spectrograph stage.

The cold stages are actively cooled by two Thales LPT9310 pulse tube cryocoolers. The internal coolant for the cryocoolers is Helium. One cryocooler is responsible for cooling the spectrograph and ACADIA. We call this cryocooler the ``100\,K cryocooler'' since its cold tip typically reaches 100\,K when the cryostat is at base temperature. The other cryocooler is solely responsible for cooling the detector; we call it the ``50\,K cryocooler'' for the same reason. Both cryocooler cold tips attach to their respective thermal stages via flexible copper-braided thermal straps (Figure~\ref{fig:cryostat_CAD}). The straps are designed to mitigate stress on the cryocoolers from thermoelastic deformations. 

\begin{figure}
    \centering
    \includesvg[width=\linewidth]{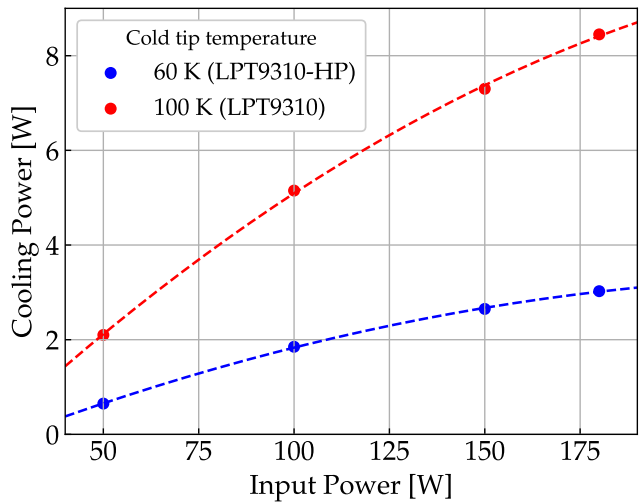}
    \caption{The performance specifications for the EXCITE cryocoolers. The data were generated with the skin temperatures kept at 20\,$^{\circ}$C and cold tip temperatures at 60\,K and 100\,K. Heat loads were applied to each cryocooler, and the input power required to maintain the cold tip at the setpoint temperature was recorded (dotted points). The dashed curves are 2nd-order fits to the data.}
    \label{fig:coolingpower}
\end{figure}

Pulse tube cryocoolers are well-suited for EXCITE. They have high technological maturity, are compact, can run indefinitely, and compared to other mechanical cooling technologies, export less vibration. However, we still take steps to reduce the impact of exported vibrations on EXCITE's pointing system. We designed the cryostat such that the compressor pistons move along the boresight. The dominant vibration direction is therefore orthogonal to the focal plane. We also use cryocooler control electronics that employ both passive and active vibration reduction algorithms.~\cite{kirkconnell2024cryocooler} Together, we have measured no negative impact from the cryocoolers on EXCITE's pointing stability. While we can measure the signature from the cryocoolers with the gyroscopes, we are not sensitive to any resultant effects on the line-of-sight stability. The cryocoolers are designed to generate a maximum cooling power of 8.5\,W at 100\,K and 3\,W at 50\,K. Thermal models of the EXCITE cryogenic receiver predict a 6.5\,W heat load with the spectrograph at 120\,K and a 0.6\,W heat load with the detector at 50\,K. Figure~\ref{fig:coolingpower} shows the cooling power curves provided by Thales of the EXCITE cryocoolers as a function of input power. 

The cryocoolers are controlled by semi-custom electronics built by West Coast Solutions (WCS). Their implementation on EXCITE is described by Kirkconnell \textit{et al.}~\cite{kirkconnell2024cryocooler} There is a separate controller for each cryocooler. The controllers operate over a wide ambient temperature and pressure range, provide active cryogenic temperature control, provide passive and active vibration reduction, are relatively lightweight, and have a small physical footprint. The controllers can lock the compressor pistons to help prevent damage during termination and landing. One cryocooler is driven at 45\,Hz and the other at 49\,Hz to avoid acoustic crosstalk that could disrupt the vibration reduction system. Accelerometers on the compressor heat sinks provide feedback for the vibration reduction system, which reduces the first, second, and third harmonics of the compressor drive frequency. Each WCS controller reads out cold thermometers inside the cryostat for active thermal control and warm thermometers on the cryocooler skins to monitor and prevent overheating.

We use a temperature controller on board the payload to read out cold thermometers located inside the cryostat. The temperature controller also supplies power to heaters installed inside the cryostat to enable active temperature control of the detector and ACADIA. The temperature controller is designed to operate in a balloon environment and is a modified version of a standard benchtop controller. We use a PC/104 computer called the Cryogenic Control Computer (CCC) to communicate between the temperature controller, the WCS controllers, and other computers on board the payload. The CCC also uses a commercial A/D board to bias and readout warm thermometers located on cryocooler skins, radiator panels, and fluid loop pumps.~\cite{kleyheeg2024integration}

\begin{figure}
    \centering
    \includesvg[width=\linewidth]{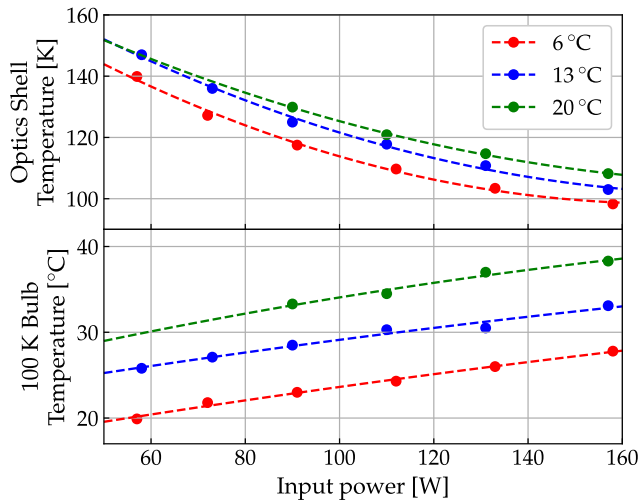}
    \caption{(Top) The temperatures of the optics shell inside the cryostat and (bottom) the external bulb temperature of the 100\,K cryocooler as a function of input power to the cryocooler and the fluid temperature set by a recirculating water chiller (6\,$^{\circ}$C, 13\,$^{\circ}$C, and 20\,$^{\circ}$C). The 100\,K bulb is more difficult to extract heat from than the compressor, which is why it tends to run the hottest and has the strongest effect on the internal temperatures.}
    \label{fig:reject_temp}
\end{figure}

The cryocoolers dissipate heat at their compressors and bulbs (see Figures~\ref{fig:cryostat_CAD} and~\ref{fig:cryostat_real}). Thermal performance studies of the LPT9310 show that heat is produced about equally between these two locations.~\cite{paine2014thermal} The heat generated increases at higher drive powers. When driven at their maximum input power, each cryocooler generates about 150\,W of heat. The cooling power of the cryocoolers is sensitive to the external skin temperatures of the compressors and bulbs and can degrade in the case of overheating. The operating temperature range for the cryocooler skins is between -40\,$^{\circ}$C and 71\,$^{\circ}$C. During lab testing, we find that the bulb of the 100\,K cryocooler runs the hottest. The cryocooler bulbs are generally more difficult to cool than the compressors because we can only make heat sinks that thermally clamp around the ``neck'' of the cryocooler, the portion which attaches the bulb and the cold finger. We quantified the relationship between the bulb temperature of the 100\,K cryocooler and the temperature of the optics shell (Figure~\ref{fig:reject_temp}). At different input powers to the 100\,K cryocooler, we measured the temperatures of the bulb and the optics shell when using a recirculating water chiller to dissipate heat from the cryocooler. We find that changes of about 1\,$^{\circ}$C to the bulb temperature corresponds to changes of about 1\,K internally. At the required operating temperature of the optics shell during flight, we expect that the bulb must remain between 45--50\,$^{\circ}$C. 

\begin{figure}
    \centering
    \includegraphics[width=\linewidth]{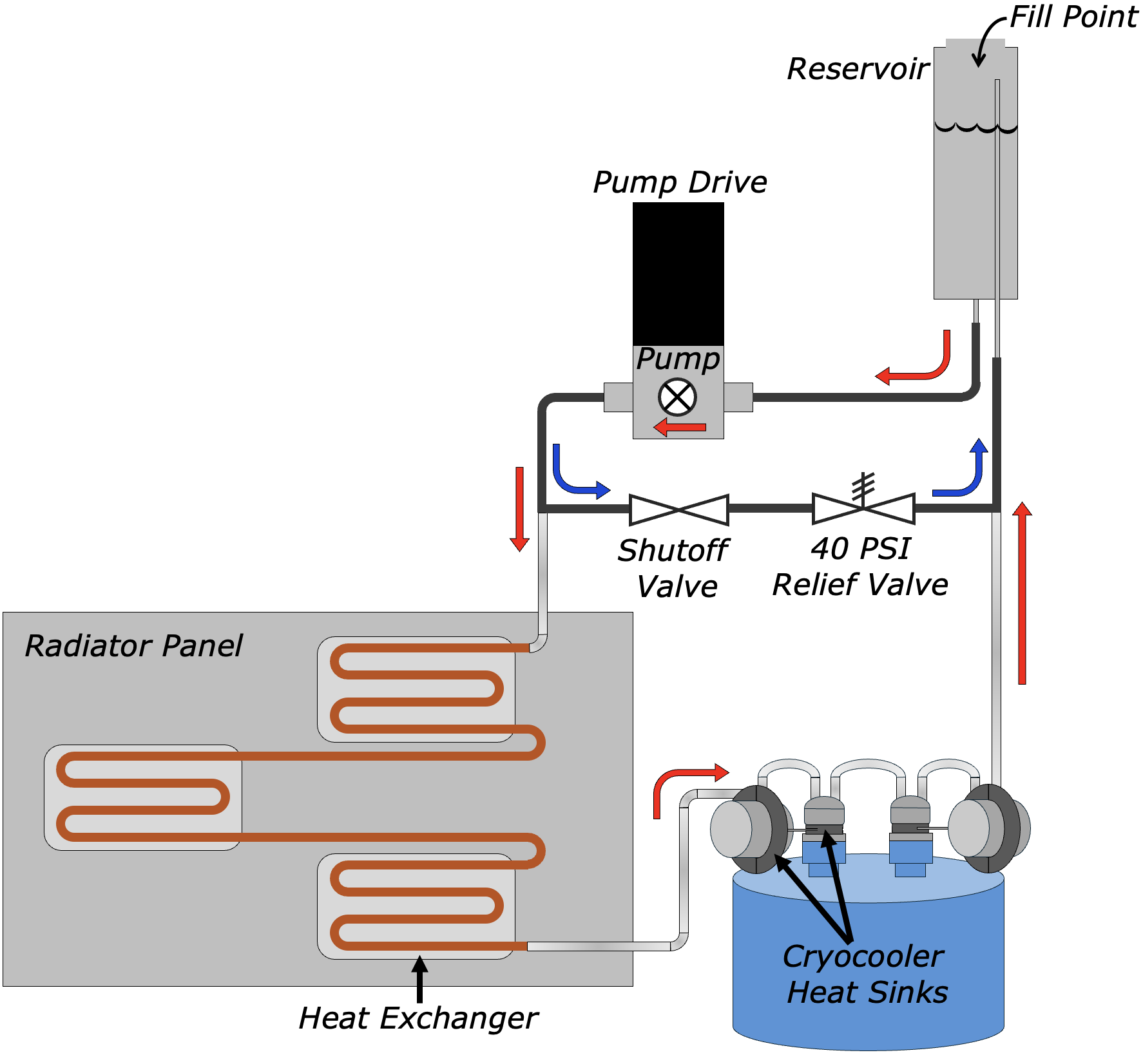}
    \caption{A schematic of a fluid loop for transporting heat from the cryocoolers to a radiator panel. The schematic shows only one fluid loop, but EXCITE flies two identical fluid loops. The reservoir is filled with methanol. The pump drives the methanol through the heat exchangers on the radiator panel, which connects to the cryocoolers and finally back to the reservoir. On the radiator panel, there are three heat exchangers embedded with copper tubing (orange). The normal flow of methanol follows the red arrows. A pressure relief loop runs in parallel with the fluid loop, and comprises a shutoff valve and a 40 PSI relief valve in series. If the pressure in the fluid loop builds up to greater than 40~PSI, then the flow of methanol will follow the blue arrows. The pressure relief loop is made from stainless steel tubing (black). Flexible tubing is used for the other parts of the loop (gray). Figure from Rehm.~\cite{rehmthesis2025} Reproduced with permission from T. Rehm, “The EXoplanet Climate Infrared TElescope (EXCITE),” Ph.D. dissertation (Brown University, 2025).}
    \label{fig:fluidloop_schematic}
\end{figure}

To dissipate the heat generated by the bulbs and compressors of both cryocoolers, we constructed two individual recirculating methanol fluid loops that transfer heat from the cryocoolers to two space-facing aluminum radiator panels placed below the bow of the gondola. We use custom heat sinks that clamp around the cryocooler compressor and necks, which are shown in Figure~\ref{fig:cryostat_real}.  Figure~\ref{fig:fluidloop_schematic} shows a schematic of this heat dissipation scheme for one of the two fluid loops. We chose methanol for its low freezing point and low viscosity across the expected operating temperature range of the fluid loops. We tested the fluid loop tubing and connections under vacuum to validate that the loop would not leak at float. The complete specifications of the fluid loop and its constituents are described by Rehm.~\cite{rehmthesis2025} Figure~\ref{fig:EXCITE_FTS} shows the radiator panels, fluid pumps, and methanol reservoirs installed on the EXCITE payload. The results from the engineering flight show that the fluid loop works thermally and does not measurably impact the pointing performance of the payload.

\begin{figure}
    \centering
    \includesvg[width=\linewidth]{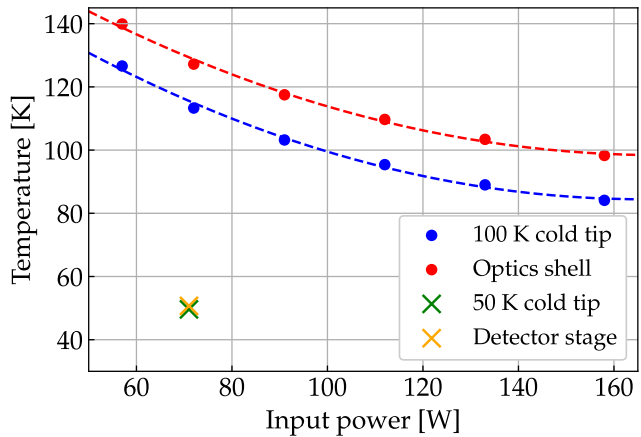}
    \caption{The temperature of the cryocooler cold tips, optics shell, and test detector stage as a function of the cryocooler input power. The optics shell temperature is a proxy for the spectrograph temperature during tests that the spectrograph is not installed. The dashed lines are second-order fits to the measured data points for the 100\,K stage. The input power to the cryocooler also creates an estimate for the heat reject requirement by the fluid loop. To get the optics shell to 120\,K, the 100\,K cryocooler would need $\sim$\,90\,W dissipated by the fluid loop. These figures agree with results from Paine.\cite{paine2014thermal} An orange X indicates the input power to the 50\,K cryocooler that results in a test detector stage cooling to $\sim$\,50\,K. The temperature of the cold tip of the 50\,K cryocooler at this power is shown as a green X.}
    \label{fig:power_100K}
\end{figure}

The thermal performance of the EXCITE cryostat was tested over a wide range of input powers, recirculating fluid temperatures, ambient pressures, and ambient temperatures. Figure~\ref{fig:power_100K} shows the results of lab tests of the measured cold tip temperature of the 100\,K cryocooler and optics shell as a function of the input power to the cryocooler. We also show the temperatures of the 50\,K cryocooler cold tip and a test detector stage. The thermal gradient across the thermal strap that connects the cold tip of the 100\,K cryocooler to the cold plate is about 14\,K at an optics shell temperature of 118\,K. Despite this gradient, EXCITE can maintain an optics shell temperature between 120--130\,K with margin, requiring only $\sim$\,100\,W of input power to the cryocooler. The thermal strap that connects the 50\,K cryocooler to the detector stage exhibited a thermal gradient of about 1\,K at a cold tip temperature of 50\,K during these lab tests. The 50\,K cryocooler cools the detector stage to 50\,K with significant margin, requiring only about 70\,W of input power. A more detailed description of the lab verification of the cryogenic system is outlined by Rehm.~\cite{rehmthesis2025}

\subsection{\label{sec:spectrograph} Spectrograph}

\begin{figure}
    \centering
    \includegraphics[width=\linewidth]{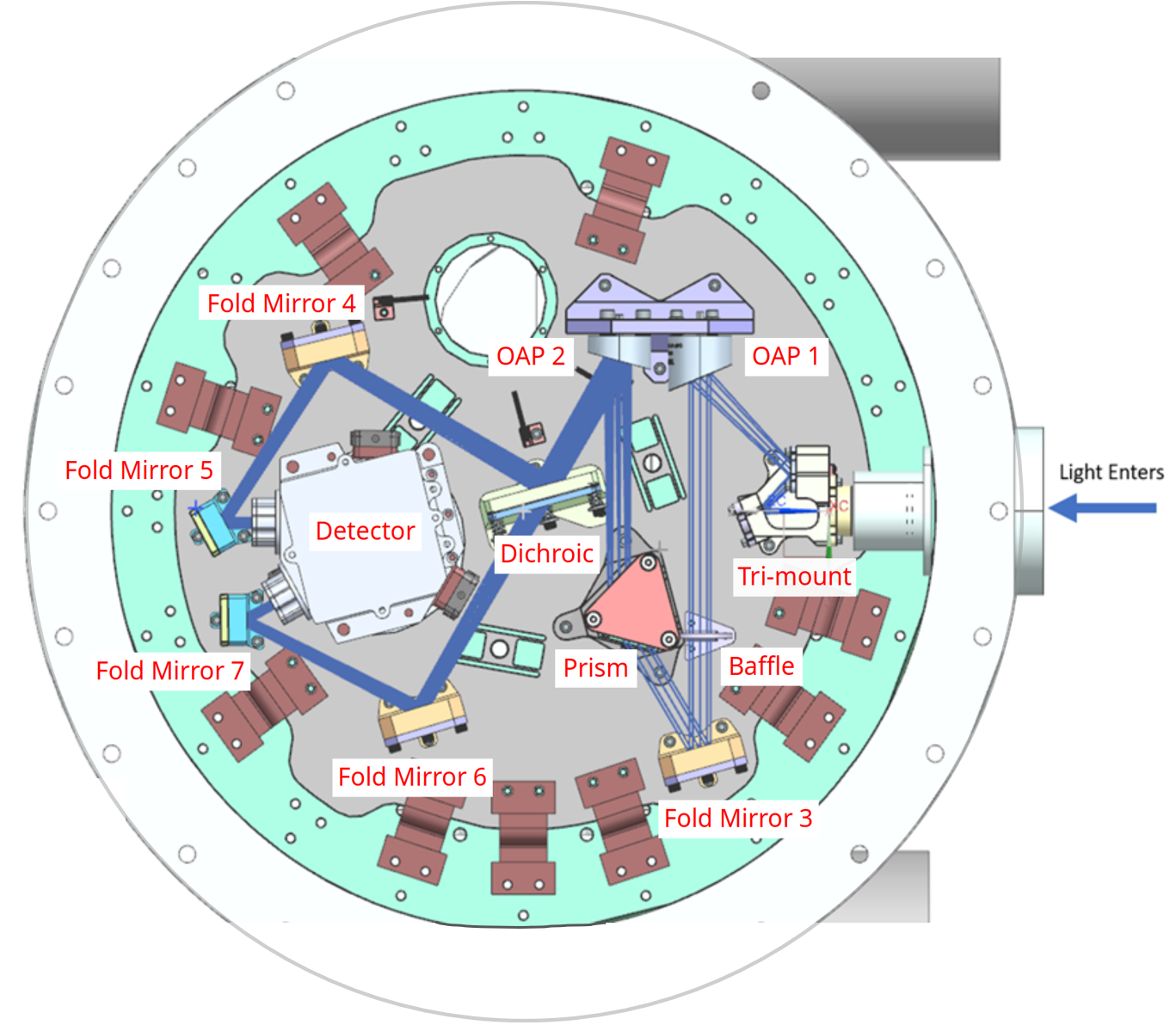}
    \caption{A top-down schematic diagram of the EXCITE spectrograph and light path.~\cite{bernard2024design} The mirror optics, field stop, and prism are commercial off-the-shelf parts, while the mounts, optics islands, and dichroic are custom-made. Reproduced with permission from Proc. of SPIE Vol. 13096, 13096A5 (2024). Copyright 2024 SPIE.}
    \label{fig:spectrograph}
\end{figure}

EXCITE employs a moderate resolution NIR spectrograph to disperse light from the target system onto the focal plane. The spectrograph is diffraction limited, which is defined as having a Strehl ratio $\geq$\,0.8. The magnification of the spectrograph is 2.7, resulting in an $f/32.2$ optical system and a detector plate scale of 230.6\,mas/pixel. Figure~\ref{fig:spectrograph} shows a CAD model of the EXCITE spectrograph with optical elements labeled. Details about the specifications of the individual optical elements and performance tolerances are provided by Bernard \textit{et al.}~\cite{bernard2022design, bernard2024design} The science beam enters the spectrograph through the cryostat window and a 100\,\textmu m-wide rectangular field stop. The width of the field stop limits stray light from entering the spectrograph and decreases background signal. The field stop is 3\,mm tall. The science beam is collimated by a 45$^{\circ}$ off-axis parabolic mirror (OAP1), redirected through a cold stop, and then dispersed by a CaF$_{2}$ prism. The dispersed beam is reflected off a 30$^{\circ}$ OAP mirror (OAP2) and then split by a cold dichroic (D2) into a transmitted (0.8--2.5\,\textmu m) component called Channel 1 (CH1), and a reflected (2.5--3.5\,\textmu m) component called Channel 2 (CH2). Both channels are then individually relayed to the focal plane by a series of two fold mirrors. The complete spectrum is imaged on about 350 detector pixels. CH1 occupies 200 spectral pixels (3.6\,mm), and CH2 occupies 150 spectral pixels (2.7\,mm).

The field stop is significantly wider than the diffraction-limited FWHM of the PSF for an $f/12$ telescope with a back focal length of 425\,nm. It has been found numerically that the FWHM~$=0.96\lambda f$, where $f$ is the focal ratio of the optical system.~\cite{bernard2022design} At 1\,\textmu m, the FWHM of the PSF equals $\sim$\,11.5\,\textmu m at the telescope focus, about a tenth of the width of the field stop. Therefore the spectrograph operates in ``slit-less'' mode, and diffraction defines EXCITE's spectral resolution limit. The predicted spectral resolution for EXCITE is $R = 80$ at 1\,\textmu m and $R \geq 50$ across the rest of the spectral range. Using the plate scale at the telescope focus and assuming an ideal Airy PSF, there is margin of about 1.5\,$''$ of pointing offset between the science beam and the center of the field stop before the FWHM is significantly clipped. In reality, this margin is more restrictive because the calculated EXCITE PSF contains a larger proportion of flux in the second Airy ring due to the relatively large secondary mirror. Regardless, we expect pointing offsets of no more than 100~mas and no significant signal loss from this effect. As shown in Section~\ref{sec:Simulations}, the field stop width is also wide enough such that we expect noise due to pointing jitter to be negligible (see Figure~\ref{fig:noisebudget}).

\begin{figure}
    \centering
    \includegraphics[width=\linewidth]{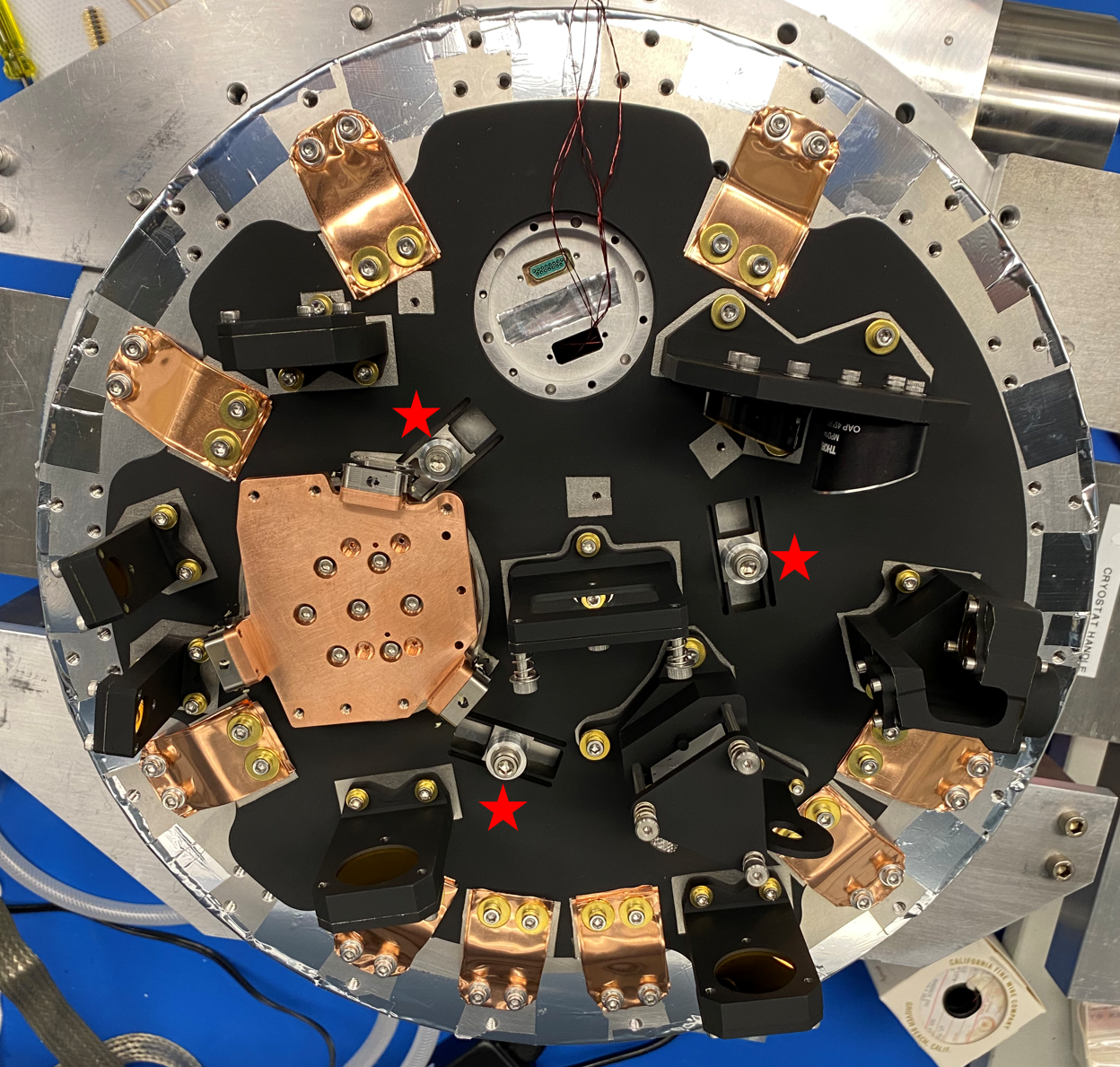}
    \caption{A photo of the EXCITE spectrograph shown in the same orientation as Figure~\ref{fig:spectrograph}. The optical mounts, islands, and bench are painted black with Aeroglaze Z306 to minimize stray light and passively cool the spectrograph. The flexure cutouts to accommodate thermal expansion mismatches are visible next to the red stars. The detector is not installed.}
    \label{fig:spectrograph_real}
\end{figure}

Figure~\ref{fig:spectrograph_real} shows a photo of the EXCITE spectrograph. The detector is not installed, but its copper cold plate which attaches directly to the 50\,K cryocooler is shown. The mounts that hold the mirrors, dichroic, and prism are fastened to ``islands'' and bolt to a single optics bench. These components are made of Ti-6Al-4V to minimize thermal deformations and stress on the optical surfaces during thermal cycling. To accommodate for coefficient of thermal expansion (CTE) mismatches between the optics bench and the aluminum cold plate inside the cryostat, three flexure cutouts are located radially about the center of the optics bench. The expected displacements are <\,300\,\textmu m across the entire optics bench when bolted to the aluminum cold plate.~\cite{bernard2024design} The optics assembly fits inside a 30.5\,cm diameter envelope and stands about 9.5\,cm tall. Nine 10-layer copper-foil thermal straps line the perimeter of the optics bench and help thermally sink the optics bench to the 100\,K stage.

The assembled spectrograph is aligned using a HeNe laser, a laser collimating optic, and an optical bench. We use the field stop entrance aperture of the spectrograph to filter the 633\,nm laser beam into a 100\,\textmu m-wide dot, approximating a single ray of light which we call the ``simulated chief ray.'' We center the simulated chief ray throughout the spectrograph optical train, which ensures the spectrograph is aligned with the optical bench. Once the spectrograph is secured to the optical bench, we remove a fold mirror and replace it with a laser collimating optic. We direct the collimated light onto OAP1, operating it in reverse such that the mirror directs light out of the field stop. We align the collimating optic by ensuring this reverse output light is maximized and horizontal relative to the optical bench. With OAP1 aligned, we reinstall the fold mirror and again recenter the simulated chief ray in all optics downstream from collimation. With the simulated chief ray and the collimated laser source aligned to the optical bench, the spectrograph is aligned to about 12\,$'$. This exceeds the alignment corresponding to 0.8 Strehl ratio (28\,$'$) by about a factor of two.

Before integrating the detector with the spectrograph, we verified the infrared throughput of the spectrograph at operating temperature using a similar reverse illumination technique. In this experiment, the infrared focal plane was replaced by what we call the ``dummy focal plane,'' a monolithic copper model of the science detector and its mounting bracket. The dummy focal plane features a small hole to allow the installation of an optical fiber at the position where we expect 1550\,nm light incident on the spectrograph to land. A jacket-less SMF-28e fiber was installed on the dummy focal plane, and it was routed out of the cryostat through a vacuum feedthrough. It was then attached to a fiber-coupled 1550\,nm benchtop laser. We then mounted an optical condenser and InGaAs photodiode outside the cryostat window to collect the light. 

The 1550\,nm light traces the optical path of the spectrograph in reverse, allowing us to verify that the spectrograph (assembled at room temperature) remains aligned during and after thermal cycling. If any optics shift, this causes the converging beam of the optical condenser to change angle and move away from the photodiode, which will show as a temperature-dependent modulation of the photodiode signal. We measured the 1550\,nm signal at the collector as a function of cryostat temperature and verified throughput down to 100\,K.

\subsection{\label{sec:detector}Detector \& Readout Electronics}

\begin{figure}
    \centering
    \includegraphics[width=\linewidth]{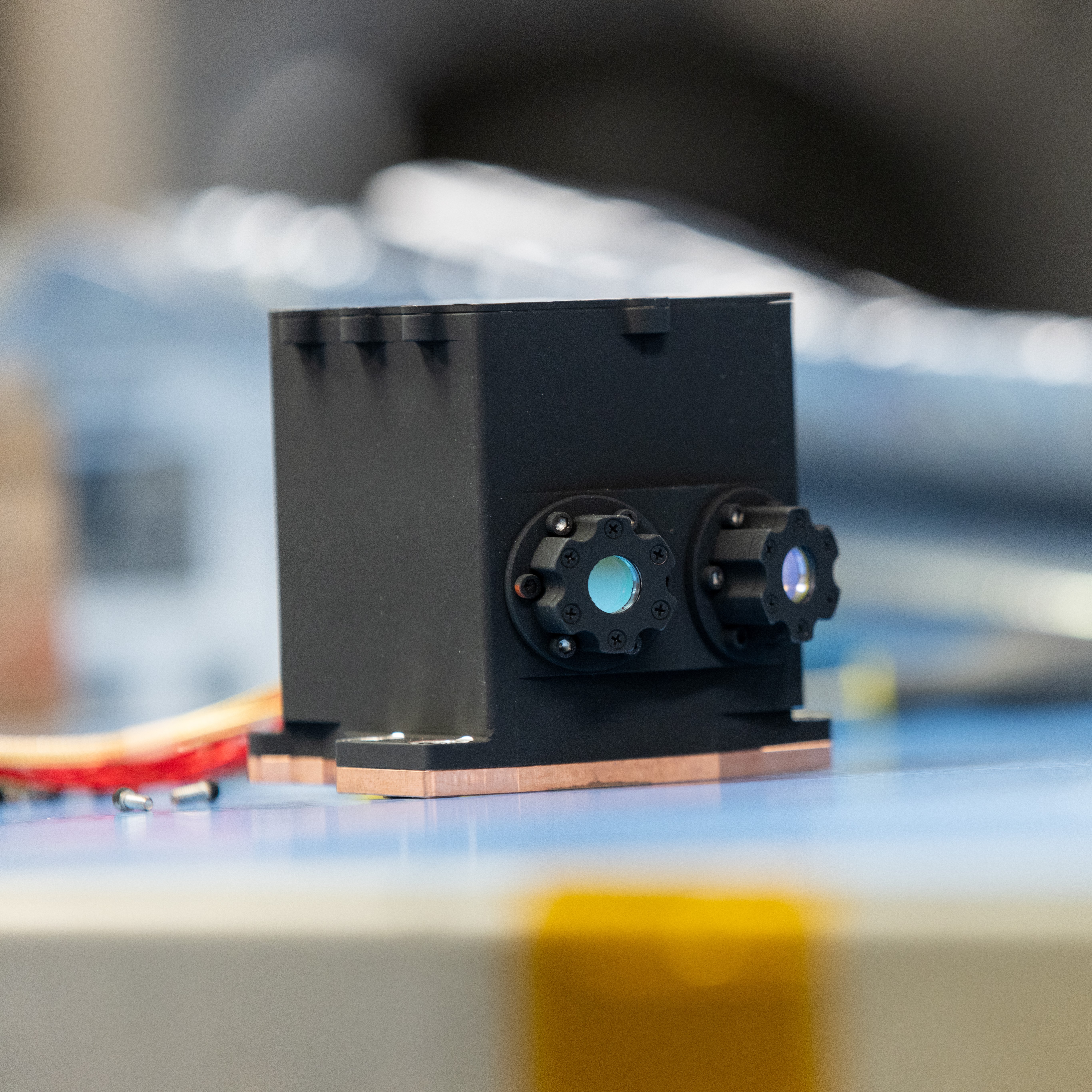}
    \caption{The EXCITE detector enclosure. The detector itself is mounted inside the black box. The two short baffle tubes allow light from each spectrometer channel to individually reach the detector. Short pass filters in the baffle tubes limit the thermal background incident on the detector. Photo taken by Sophia Roberts/NASA.}
    \label{fig:EXCITEdetector}
\end{figure}
The EXCITE infrared detector is a Teledyne HgCdTe Astronomy Wide Area Infrared Imager with 2k resolution, reference pixels, and guide mode operation (H2RG).~\cite{loose2003hawaii} The detector used by EXCITE during its test flight was an engineering unit left over from \textit{JWST}/NIRSpec development. For the LDB science flight, EXCITE will use a NIRSpec flight candidate H2RG whose performance is summarized by~\citet{rauscher2014new}. The detector is sensitive between 0.8--5.3\,\textmu m. For operating temperatures below $\sim$\,65\,K, the detector's dark current noise contribution to EXCITE's error budget is negligible.~\cite{rauscher2011dark} Compared to other NIR detector technologies (\textit{e.g.}, InSb), HgCdTe detectors achieve lower dark current at higher temperatures, matching EXCITE's cryogenic solution. The H2RG is located in a blackened box that has two apertures to allow incident light to shine from each of the spectrograph channels. These apertures, located at the end of short baffle tubes, contain short-pass filters with a cutoff wavelength of 3.5\,\textmu m. The filters block unwanted thermal emission from the optics, and the baffle configuration limits stray light from the spectrograph. A photo of the EXCITE detector enclosure is shown in Figure~\ref{fig:EXCITEdetector}. The detector has four output channels, each corresponding to a 2048\,$\times$\,512 pixel quadrant of the detector. The spectra land on channels 1 and 4. Channels 2 and 3 sample stray light. The maximum flux per pixel on the H2RG for a bright target is 5863\,e$^{-}$/s in CH1 and 1283\,e$^{-}$/s in CH2, corresponding to saturation times of about 20.5\,s and 93.5\,s, respectively.

The detector is read out using an ACADIA controller.~\cite{loose2018acadia} The ACADIA was developed for the \textit{Roman Space Telescope} (\textit{RST}). The ACADIA is maintained at a stable temperature near 130\,K. The intended detector readout scheme is described by~\citet{nagler2022exoplanet} and is further detailed in Section \ref{sec:Simulations} of this paper. The ACADIA and H2RG are read out by a Multi-purpose ASIC Control and Interface Electronics (MACIE) controller card. The MACIE is designed to read out Teledyne HxRGs. The MACIE operates at ambient temperature and is installed inside the Detector Control Box (DCB), located on the underside of the gondola's inner frame.

The EXCITE science detector was extensively characterized by the NIRSpec team, with performance and operation verified by EXCITE. Custom firmware was developed to interface the ACADIA with EXCITE's detectors. It has been validated in both laboratory and flight settings. Assuming that we will read two windows of 512\,$\times$\,248 pixels (see Section \ref{sec:Simulations}), the detector system's data rate is 350~MB per hour in flight. A 60-day flight would result in just over 500~GB of detector data. 

\section{\label{sec:Simulations}Simulation Tools, Data Reduction, \& Systematics}
We have developed simulation tools to quantify EXCITE's expected sensitivity and evaluate the impact of systematic effects we expect to encounter with EXCITE. These include both correlated and uncorrelated effects. Building upon work by Nagler \textit{et al.},~\cite{nagler2019observing} we show that we expect EXCITE to be background-limited across its entire band. We have also quantified noise due to pointing jitter, evaluating the impact of both absolute pointing errors (APE) and random pointing errors (RPE). We find that noise due to jitter is subdominant to radiometric noise. We have also generated time series of simulated focal plane images that contain full-orbit phase curves of hot Jupiters.

To generate the time series spectral images and perform radiometric calculations, we use the~ExoSim2 simulator.~\cite{Mugnai_2025, mugnai2024exosim, mugnai2023exorad} ExoSim2 is a highly-configurable time domain end-to-end simulator that also provides radiometric signal and noise estimates for instruments intended to perform point source photometry and spectroscopy. ExoSim2 can generate spectral images that may include photon noise, pointing jitter, detector dark current, read noise, and more. We use the dynamic capabilities of ExoSim2 to create stacks of spectral images and inject the exoplanet signal as well as balloon systematics in order to perform statistical detrending of the phase curve amplitude.~\cite{rehmthesis2025}
Section~\ref{sec:radiometric} describes the methods used to calculate noise from uncorrelated sources and shows a radiometric model for EXCITE. Section~\ref{sec:jitter} details the expected noise contributions due to jitter and reports a total noise budget for EXCITE. Section~\ref{sec:ELDiP} describes the front-end data reduction pipeline that we plan to use during flight in order to monitor data quality at the detector stage and assess real-time instrument performance.

\subsection{\label{sec:radiometric} Uncorrelated Noise}
\begin{figure}
    \centering
    \includesvg[width=\linewidth]{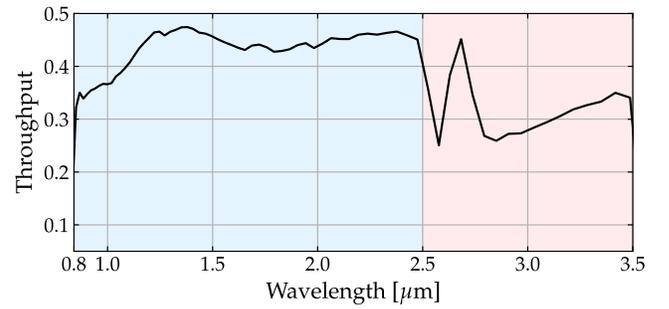}
    \caption{The total throughput of the EXCITE optical system binned to $R=50$. The two science channels are split by a dichroic at 2.5\,\textmu m. This band edge was chosen based on noise calculations. Beyond $\sim$\,2.5\,\textmu m, the noise budget is dominated by the thermal emission from ambient temperature optics. Shorter than $\sim$\,2.5\,\textmu m, the noise is dominated by photon noise from a the host star. The transmission profile of the cold dichroic (D2) at an angle of incidence of 45$^{\circ}$ contains a peak at 2.68\,\textmu m and dips at 2.85\,\textmu m, which accounts for the features in the spectrum beyond 2.5\,\textmu m. The blue shaded region outlines the spectral coverage of CH1, and the red shaded region outlines the spectral coverage of CH2.}
    \label{fig:throughput}
\end{figure}

To quantify the photon noise from uncorrelated sources, we calculate the photon signal from the target star, the Zodiacal background, the foreground emission from the Earth's atmosphere, and the thermal emission from the instrument. Photon signals are assumed to obey Poisson statistics. We account for the total optical efficiency of the instrument using measured and/or calculated values of each optical element's transmission or reflection spectrum, and we account for the detector's quantum efficiency using data presented by Rauscher \textit{et al.}~\cite{rauscher2014new} The total throughput of the EXCITE optical system is shown in Figure~\ref{fig:throughput}. Once the total throughput and photon signals are calculated, ExoSim2 uses an empirically found wavelength solution for the EXCITE optical system and estimates synthetic apertures for each spectral bin, where we set the encircled energy to be 91\%. The radiometric output of the simulator is a table containing signals and noises in units ct/s per spectral bin.

\begin{figure}
\centering
\includesvg[width=\linewidth]{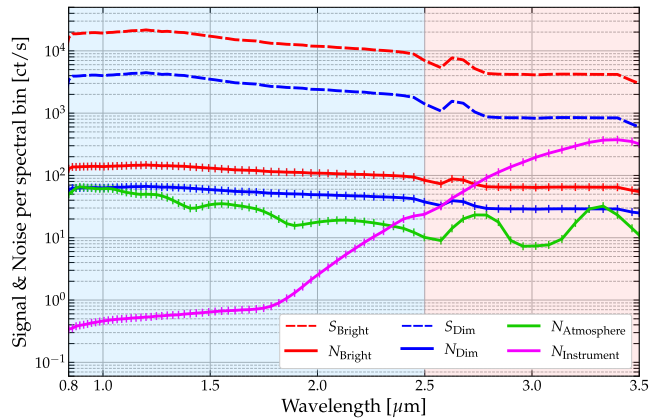}
\caption{\label{fig:radiometric_model} The radiometric model for EXCITE, generated using ExoSim2. The dashed lines show the signal levels for typical bright and dim targets ($S_{\text{Bright}}$, $S_{\text{Dim}}$) that EXCITE can observe from the Antarctic. The solid lines are noise profiles for the targets ($N_{\text{Bright}}$, $N_{\text{Dim}}$) as well as the Earth's atmosphere ($N_{\text{Atmosphere}}$) and the instrument emission ($N_{\text{Instrument}}$). The noise profiles are scaled to 1\,s integration times and scale down with $\sqrt{t_{\text{int}}}$. The vertical ticks on the noise profiles indicate the central wavelength for each spectral bin at $R=50$. Only results of modeled photon noise sources are shown, as the noise contribution from jitter is handled differently (see Section~\ref{sec:jitter}). The blue shaded region outlines the spectral coverage of CH1, and the red shaded region outlines the spectral coverage of CH2.}
\end{figure}

Figure~\ref{fig:radiometric_model} shows a radiometric model for EXCITE for a bright (8.13~K$_{s}$-mag) and dim (9.37~K$_{s}$-mag) target visible during an Antarctic LDB flight. The results are binned to $R = 50$ and include the noise from the target stars, the Earth's atmosphere, and instrumental thermal emission. The spectral radiance of the Earth's atmosphere is calculated using MODTRAN~\cite{berk1999modtran4} and assumes 38\,km altitude, 45$^{\circ}$ elevation, 55$^{\circ}$ solar elevation, and 180$^{\circ}$ anti-Sun azimuth. The ambient temperature optics, \textit{i.e.}, the primary mirror, secondary mirror, tip/tilt stage, D1, and cryostat window, are set to 273\,K. The ``cold optics,'' \textit{i.e.}, the optical elements and shrouds inside the cryostat, are set to 120\,K. The detector enclosure is set to 50\,K. As shown, EXCITE is photon-noise limited by the target star between~0.8--2.73\,\textmu m for a bright target and between~0.8--2.63\,\textmu m for a dim target. At longer wavelengths, the noise is dominated by thermal emission from the ambient temperature optics. The cold optics do not contribute significantly to the noise as long as they are cooled to $T$\,<\,160\,K. At wavelengths <\,1\,\textmu m, for a dim target, noise from Earth's atmosphere is comparable to noise from the target star.

\subsection{\label{sec:jitter}Pointing Jitter}
Pointing jitter along the telescope boresight is another source of noise. It leads to variable losses at the field stop and signal modulation at the science focal plane. We treat these two situations separately. We show that even in the absence of advanced detrending techniques, noise due to jitter is always subdominant to photon noise. In principle, the spectral image gives a precise measurement of the line-of-sight, and if the throughput loss as a function of line-of-sight variations at the field stop is known, then the noise induced by jitter can be completely detrended. Recent work by~\citet{bocchieri2025jittering} outlines a method for detrending jitter from spectral images generated by ExoSim2. They find that, even in the case when $N_{\text{jitter}}$ is much less than other sources of photon noise, properly detrending jitter significantly decreases the bias of statistically retrieved transit depths.

To estimate the noise associated with pointing jitter at the field stop ($N_{\text{field stop}}$), we first calculate the EXCITE PSF using a 2D Fourier transform of the as-built telescope aperture. We then calculate the width of the field stop in the image plane for each wavelength and generate an RPE timestream that moves the image of the field stop on the PSF. EXCITE uses a rectangular field stop with a width of 100\,\textmu m; the plate scale of the optical system at the field stop is 34.4\,mas/\textmu m. At each time step, we integrate the flux, $F$, of the PSF inside the field stop. The resulting noise is calculated as $N_{\text{field stop}} = \text{std}(F)/\bar{F}$ over the entire timestream, where $\text{std}(F)$ is the standard deviation of the flux and $\bar{F}$ is the average flux. We can vary the APE (offsets from the beam center to the field stop), RPE, and integration time. Based on the pointing results from the 2023 SuperBIT flight,\cite{gill2024superbit} we created our timestreams with a 1-$\sigma$ RPE of 50~mas at 1\,Hz, APE of 100 mas, and 2\,hr integration times. We find that $N_{\text{field stop}} <25$\,ppm across 0.8--3.5\,\textmu m. $N_{\text{field stop}}$ peaks at around 2.3\,\textmu m. At this wavelength, the field stop starts to vignette the first sidelobe of the PSF. Similarly, the jitter noise is relatively large at 0.8\,\textmu m and 3.4\,\textmu m due to vignetting of the second sidelobe and the main lobe, respectively.

To simulate noise due to jitter at the science focal plane ($N_{\text{focal plane}}$), we centered an Airy profile with diameter $D(\lambda)$ on a grid of detector pixels where the QE of each pixel was only known to $\pm$0.1\% at 1-$\sigma$. Using a plate scale of 12.8\,mas/\textmu m at the focal plane, we created another jitter timestream with 1-$\sigma$ RPE of 50~mas at 1\,Hz. At each point in time, we jittered the PSF location and integrated the flux of the PSF multiplied by a random QE map with the assumed uncertainty. We used rectangular apertures that have widths of the nearest integer number of pixels such that $\sim$\,$2.5D(\lambda)$ were captured, or about $91\%$ of the encircled energy of the PSF. The apertures were centered on the middle of the grid and remained static through time since 50\,mas corresponds to a linear translation of about one-fifth the size of a pixel on the focal plane. The resulting noise $N_{\text{focal plane}}$ is calculated in the same way as $N_{\text{field stop}}$. We find that $N_{\text{focal plane}} <24$ ppm at 2\,hr integration times. $N_{\text{focal plane}}$ peaks at 0.8\,\textmu m where the PSF is concentrated on the fewest number of pixels. It decreases like $1/\lambda$ out to 3.5\,\textmu m. The total noise due to jitter is given by $N_{\text{jitter}}=\sqrt{N_{\text{field stop}}^{2}+N_{\text{focal plane}}^{2}}$ and is shown in Figure~\ref{fig:noisebudget}. The calculated jitter noise is subdominant across the EXCITE band.

Figure~\ref{fig:noisebudget} shows the sensitivity in ppm for EXCITE when combining radiometric sources and pointing jitter. We show the contributions to the total sensitivity from the target star, pointing jitter, and other radiometric sources, \textit{i.e.}, the Earth's atmosphere and instrument emission. The sensitivity profiles are plotted as $N/S_{\star}$, where $N$ is the noise from a source or a combination of sources, and $S_{\star}$ is the signal from the target star. The results are scaled to integration times of 2\,hr, which is about the duration of the secondary eclipse for many of EXCITE's candidate targets. The secondary eclipse depths are calculated using Equation~\ref{eq:depths}.

\begin{figure}
    \centering
    \includesvg[width=\linewidth]{Figures/sensitivity_total_jitter.svg}
    \caption{The sensitivity per spectral bin for EXCITE. The dashed lines show the contribution to the sensitivity from the photon noise of a bright and dim target. The dotted lines show the combined contribution to the sensitivity from radiometric sources (Rad), including the Earth's atmosphere and instrument emission. The solid orange line shows the contribution to the sensitivity from pointing jitter. The total sensitivity for both targets are shown as solid lines. The solid lines with vertical ticks show the secondary eclipse depths calculated for each  target ($D_{e}$). The vertical ticks indicate the central wavelength for each spectral bin. The sensitivity is scaled to 2\,hr integration times, about the duration of the secondary eclipse for these test targets. EXCITE is photon-noise limited by the target for all of CH1. In CH2, the Earth's residual atmosphere and ambient optics dominate the total signal. At this integration time, the eclipse depths are above the sensitivity across the entire EXCITE passband for a bright target and from 1--3\,\textmu m for a dim target. The blue shaded region outlines the spectral coverage of CH1, and the red shaded region outlines the spectral coverage of CH2.}
    \label{fig:noisebudget}
\end{figure}

\begin{figure}
    \centering
    \includesvg[width=\linewidth]{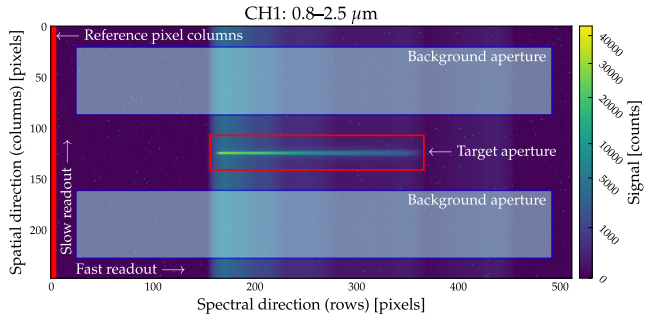}
    \caption{A simulated spectral image of a bright target in CH1. This image was produced using ExoSim2 and shows a non-destructive read (NDR) on a 512\,$\times$\,248 pixel grid after about 4\,s of exposure time. The $y$-axis is the spatial direction and the $x$-axis is the spectral direction. The target spectrum is contained within the target aperture outlined in red. The rectangular regions in light blue with dark blue borders outline the pixels that are used for background subtraction. The solid red rectangle on the left-hand side of the image shows the columns that are used as reference pixels in CH1. The different directions for fast and slow readout are also shown. This image includes dispersed foreground signal from the Earth's atmosphere and instrument emission, which can be seen spanning the entire spatial direction from pixel 150 to pixel 460 on the spectral axis. It also includes dark current, read noise, QE variations, hot pixels, kTC noise, and pixel nonlinearities. On the actual 2048\,$\times$\,2048 focal plane, the spectral image of CH1 is located on the leftmost quadrant of the detector (quadrant 1), and CH2 is located on the rightmost quadrant (quadrant 4).}
    \label{fig:spectralimage}
\end{figure}

\subsection{\label{sec:ELDiP}Simulated Spectral Images \& Line-of-Sight Pipeline}

\begin{figure}
    \centering
    \includegraphics[width=\linewidth]{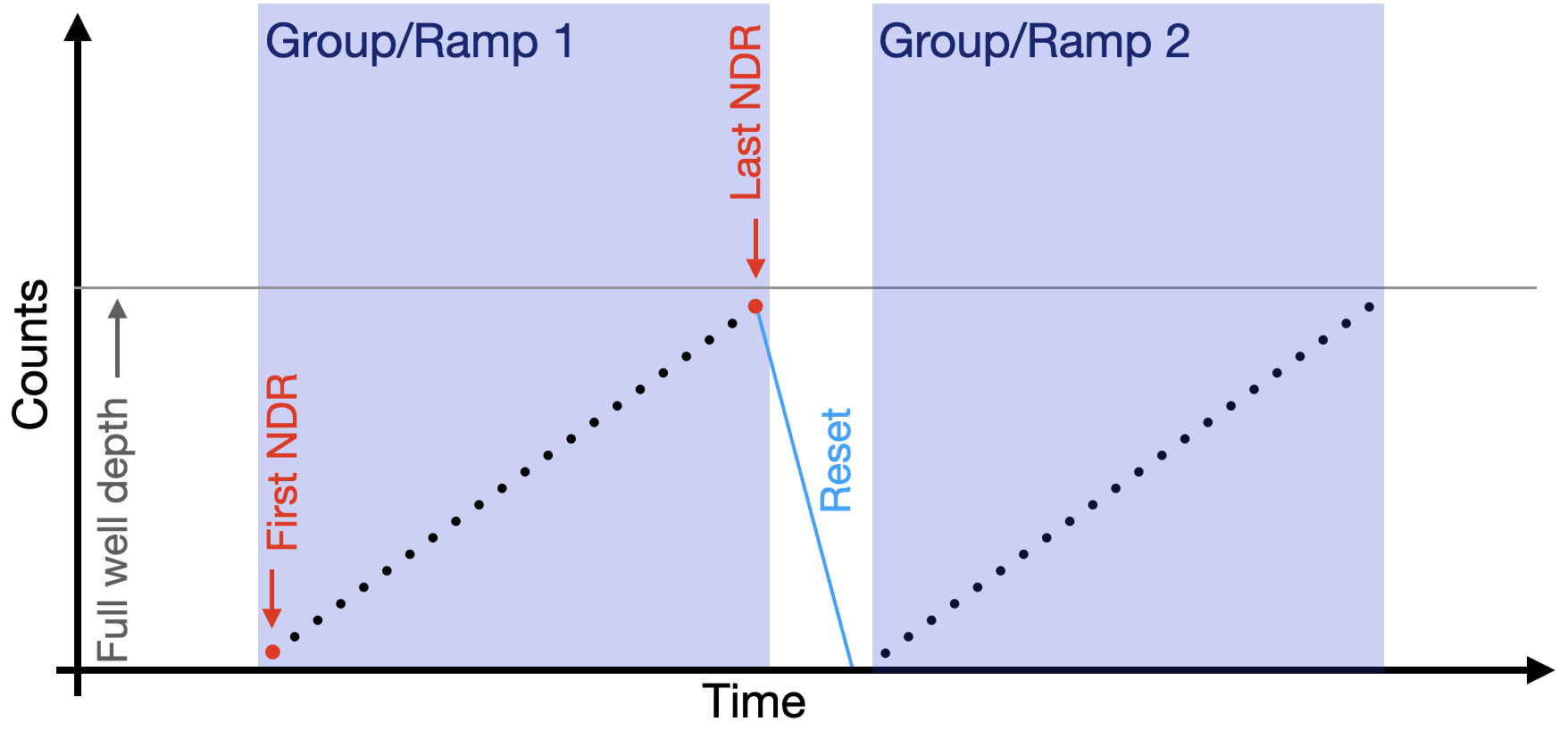}
    \caption{A diagram of the readout scheme of EXCITE. Each EXCITE ``observation'' consists of multiple ``groups'' of NDRs, which are discrete detector measurements taken without a detector reset. In early stages of data processing, groups may also be referred to as ``ramps'' due to the build-up of charge on a detector pixel. Once a ramp approaches the full-well depth (the maximum charge a pixel can store before saturation), it is reset, and a new ramp begins after a chosen amount of time. For simplicity, the ramps are drawn as linear.}
    \label{fig:ramp-readout}
\end{figure}

When the payload is launched from the Antarctic, there will be a period of time when we maintain line-of-sight contact with the instrument, enabling high-rate telemetry. During this period we will tune the instrument for the float environment and verify that the instrument is performing nominally at the beginning of the flight. We call this the ``line-of-sight'' portion of the flight. The EXCITE Line-of-sight Diagnostic Pipeline (ELDiP) is a data reduction pipeline with multiple diagnostic features. The ELDiP is designed to quickly evaluate data from the detector system as it is downloaded during this early stage of the flight, check for nominal detector and optical behavior, and serve as a preliminary quick-look data reduction and analysis tool for observed spectra. Simulated data using ExoSim2 were used to validate this pipeline (see Figure~\ref{fig:spectralimage}), with a data structure mimicking that generated by instrument's detector system. The pipeline is designed to be used with observed data but has thus far only been tested with simulated data. The terminology used to describe different aspects of the detector readout is described by Figures~\ref{fig:spectralimage} and \ref{fig:ramp-readout}.

The ELDiP operates in several consecutive but independent stages:

\textbf{Stage 0} consists of data read-in and set-up. This stage is performed on one group of non-destructive reads (NDRs) at a time (\textit{i.e.}, the input will consist of several consecutive NDRs forming one group/ramp). An NDR is a discrete voltage measurement across a window of detector pixels to measure signal accumulation without resetting the stored charge in each pixel. Data is separated into its two separate science channels, and several observing parameters are set: the exposure time of each NDR, detector well depth, gain, and read noise. Stage 0 outputs a separate \texttt{FITS} file for each science channel, which are used in the subsequent pipeline steps.

\textbf{Stage 1} consists of NDR corrections which do not depend on the observed target. Steps of this stage are performed on the outputs of Stage 0, \textit{i.e.}, \texttt{FITS} files containing groups of NDRs, either from CH1 or CH2. 
We first perform a saturation correction. Saturated pixels are identified as any pixel exceeding the full well depth (shown in Figure~\ref{fig:ramp-readout}). Saturated pixels in the dataset are masked, and their values are not considered in future steps.  
A check is performed to ensure that any pixel flagged as saturated is also flagged in subsequent NDRs within the group. 
An optional step also flags the 4 fully bordering pixels of any saturated pixel as also saturated. 
After this step, we perform a dark current subtraction. We assign each detector pixel a dark current value based on dark exposures performed prior to flight. The dark current pixel map is subtracted from each NDR. 
Then, we perform a reference pixel correction. Reference pixels are defined as the 4 leftmost columns for CH1 and the 4 rightmost columns for CH2. Here, each ``column'' spans the spatial direction, and each ``row'' spans the spectral direction. As shown in Figure~\ref{fig:spectralimage}, the CH1 ``fast'' readout direction is left-to-right, such that the reference pixels are read out at the beginning of each row. For CH2, fast readout is performed right-to-left, such that the reference pixels in the rightmost column are read out at the beginning of each row. The ``slow'' readout direction is then vertical: each entire row is read out in the fast scan direction before moving up column-wise in the slow direction. Following Bushouse \textit{et al.},~\cite{jwst_pipeline} we use a sliding average method with a tunable box size to determine the reference pixel correction for each row of data. The reference pixel values are subtracted from each row in each NDR. 

A nonlinearity correction is performed on the group to account for systematic nonlinearities in the charge accumulation of the detector pixels. Systematic nonlinearities are first characterized from flat-field images taken prior to science flight. For benchmarking, we used simulated flat-field groups with 5\% deviations from linear to test the functionality of this step in the pipeline. We characterize the detector pixel nonlinearity and apply the correction to our data based on the method described by Hilbert.~\cite{hilbert2014} Finally, we identify and mask out pixels containing data jumps between consecutive NDRs, such as cosmic rays and dark pixels. The algorithm used to define a data jump is based on the algorithm described by Anderson and Gordon.~\cite{jumpmask2011} This step of the pipeline iterates over the algorithm, identifying and flagging outlier pixels until no more jumps are detected.

\textbf{Stage 2} performs corrections which are dependent on the location of the target spectrum on the detector. Within each NDR, we first place a large rectangular aperture around the entire spectral trace of the observed target. To identify the spectral boundaries of the aperture, we take column-wise medians of the NDR and difference adjacent column medians to locate jumps at the start and end of the spectral trace. To identify the aperture's spatial boundaries, we take row-wise medians of the NDR and fit a Gaussian to the row medians to locate the central pixel of the target PSF. The spatial boundaries are then defined as the nearest integer number of pixels that encompasses the FWHM of the PSF.

After the rectangular target aperture is established, the background is subtracted to remove $1/f$ noise. We define rectangular background apertures as a default 20 pixels away from the edges of the detector in all directions and 20 pixels spatially separated from the defined target aperture. The background values are defined as the median value in each column within the background apertures, and a column-wise background subtraction is performed over the entire NDR. Locations of target and background apertures are shown in Figure~\ref{fig:spectralimage}.

For an instrument performance check, the target trace is then split up into spectral bins given the spectral resolution of the spectrograph. The same method of finding the spatial aperture boundaries is then applied to each spectral bin, allowing the spatial aperture size to vary over wavelength depending on the width of the PSF. At each spectral bin, the target trace centroid, width, and symmetry versus wavelength is plotted to ensure the trace is optimally positioned on the detector and that optical performance is nominal. If this stage identifies variations in PSF shape or large changes in the position of the spectral trace, optical alignment can be checked and motors can be tuned while the instrument is within line-of-sight.   

After all NDR-level corrections have been performed, a ramp-fitting step determines the mean count rate for each pixel within each group. This step performs a linear fit to each pixel ramp (nonlinearity corrects have already been applied), with custom treatment to avoid masked pixels--data jumps or saturated pixels--within the group. At the end of Stage 2, all NDR-level and group-level corrections have been performed. Each group is represented by one count rate (flux) value for further time-series analysis. 

\textbf{Stage 3} creates a time series of the resulting flux values after Stages 0--2 have been performed on each group. Stage 3 creates a singular image using the time-median flux in order to check the target spectrum. The spectral bins defined in Stage 2 are used to create a spectrum of the target, which is compared against a spectral model to monitor data quality. The ELDiP is programmed to compare the observed spectrum against a blackbody spectrum of the target's temperature, but can receive more specific input spectral models. This stage also produces white light curves for each of the science channels. During the line-of-sight portion of the flight, we do not expect to observe transits. In the case we do, Stage 3 has an optional step to fit white light curves to \texttt{batman}~\citep{batman2015} transit models. All further data reduction and analysis will be performed by more specialized pipelines.

When the ELDiP is run front-to-back, it has an option to produce plots for visual checks of each correction step. It will automatically save a flag for 1) temporal or spectral variations of the trace centroid location; 2) PSF asymmetries; and 3) large differences in observed versus model spectra. In doing so, the pipeline is designed to identify issues that are correctable either at the detector stage, with the pointing system, or in the optical alignment early in the flight and before any science targets are observed.

\section{\label{sec:FTS}Integration and Testing and Flight Campaigns}

In preparation for the flight campaigns, the EXCITE payload was tested extensively at the subsystem level. This included 1) initial on-sky performance verification of the gondola pointing system at StarSpec Technologies; 2) as-delivered telescope alignment verification; 3) the first complete telescope and gondola assembly at Brown University; 4) telescope alignment and on-sky pointing verification at CSBF in Palestine, Texas; 5) thermal vacuum (TVAC) testing of the gimbal motors and flight computers at CSBF in Palestine, Texas; 6) TVAC testing of the cryostat, fluid loop, and WCS controllers at NASA Goddard Space Flight Center (GSFC); and 7) extensive testing of the complete EXCITE instrument before flight at CSBF in Fort Sumner, New Mexico. Section~\ref{sec:integration} outlines the integration procedure of the EXCITE payload after component-level testing. Section~\ref{sec:2023} describes the results from the 2023 Fort Sumner campaign. Section~\ref{sec:2024} discusses the performance of EXCITE during its engineering flight from the 2024 Fort Sumner campaign. The purpose of the engineering flight was to validate instrument operation before an LDB mission from the Antarctic. Much of the gondola/ACS has flight heritage, but we demonstrated the complete science instrument for the first time from a balloon platform. Section~\ref{sec:refurb} details the refurbishments and improvements made to the payload since the engineering flight, in preparation for an Antarctic LDB flight.

\subsection{\label{sec:integration}Payload Integration}

After lab and TVAC testing to verify component-level functionality, we prepare the payload for flight using the following top-level integration procedure:
\begin{enumerate}
    \item Install the ACADIA inside the cryostat;
    \item Install the spectrograph inside the cryostat;
    \item Install the detector inside the cryostat;
    \item Close cryostat and verify detector system operation;
    \item Integrate telescope with inner frame;
    \item Align the primary and secondary telescope mirrors using a Takahashi collimating scope;
    \item Align a collimated beam source to the telescope primary mirror input (a so-called \textit{artificial star});
    \item Using the artificial star, align the visible light telescope beam with FGC by adjusting the angle of the tip-tilt mirror mount;
    \item Integrate the cryogenic receiver with the telescope and inner frame;
    \item Using the artificial star and the FSVC, align the infrared telescope beam with the entrance field stop of the spectrograph by adjusting tip-tilt on D1;
    \item Mount the middle frame inside the outer frame of the gondola;
    \item Integrate the science instrument with the gondola by mounting the inner frame on the middle frame;
    \item Put on the gondola hat by attaching to the outer frame;
    \item Cool the cryostat and tune the cryogenic system's vibration reduction system; 
    \item Complete integrated system tests, including on-sky optical and pointing verification tests;
    \item Integrate payload with CSBF instrument package and verify compatibility between the two systems.
\end{enumerate}
Once compatibility with CSBF hardware is confirmed, the payload may be declared flight-ready.

\subsection{\label{sec:2023}2023 Fort Sumner Campaign}
The 2023 balloon campaign was EXCITE's first comprehensive I\&T campaign. We declared flight readiness in early September, but were unable to launch due to poor weather and operational priorities over the next month. Nonetheless, this campaign included extensive ground testing and many nights of on-sky operations. We verified successful ground operation of the whole instrument, as well as compatibility with CSBF hardware.

EXCITE performed its first fully-integrated on-sky pointing tests during the 2023 campaign, including 3-axes pointing with tracking on both star cameras. Figure~\ref{fig:EXCITE_FTS} shows the EXCITE payload during an on-sky pointing test from the highbay in Fort Sumner. We showed seeing-limited ground pointing down to 1--3\,$''$ (1-$\sigma$). During one ground-pointing session, the boresight star camera stabilized with 1-$\sigma$ pointing of 2.63\,$''$ and 1.67\,$''$ in the cross-elevation and elevation directions, respectively. The cross-elevation direction is orthogonal to the elevation axis. With the entire instrument  operational, the ACS performed nominally. Moving parts from the cryogenic system did not measurably impact pointing performance. Based on the pointing stability of the EXCITE payload during ground-pointing and comparing with the in-flight performance of the SuperBIT experiment, we expect gondola stability at the 0.2--0.3\,$''$ (1-$\sigma$) level for EXCITE.

\begin{figure}
    \centering
    \includegraphics[width=\linewidth]{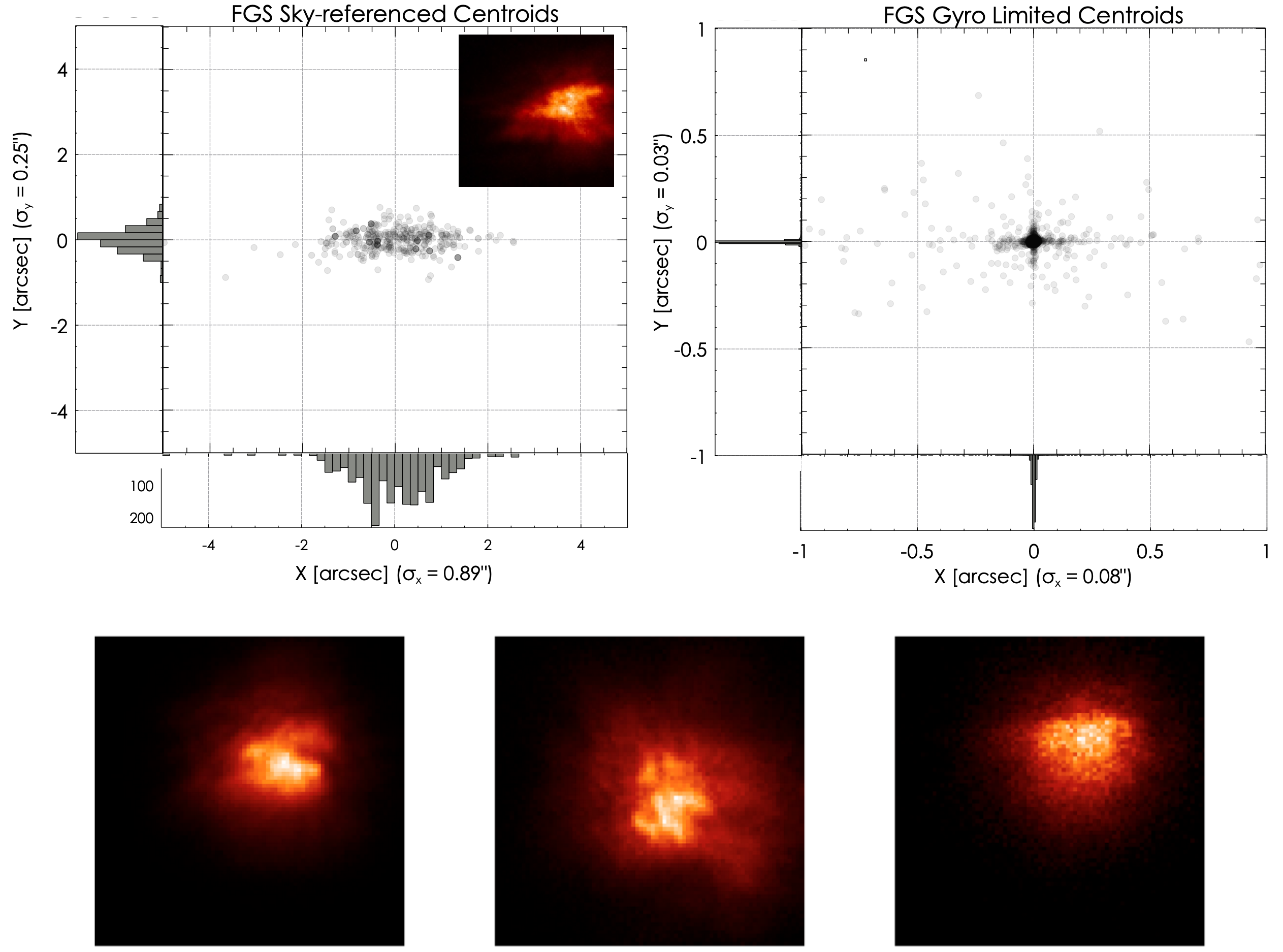}
    \caption{Pointing performance of the FGS from the ground during the 2023 Fort Sumner campaign; on-sky (top-left) and gyroscope-only (top-right) FGC centroid distributions over 5~minutes of observation are shown (black) with 2D distributions per axis (gray). Sample thumbnail images of the target star during this demonstration run are shown (bottom row) to emphasize the effects of ground-based seeing in obfuscating pointing feedback. Figure from~\citet{romualdez2024exoplanet} Reproduced with permission from Proc. of SPIE Vol. 13094, 130944Y (2024). Copyright 2024 SPIE.}
    \label{fig:FGS2023}
\end{figure}

In addition to stabilizing the gondola, we also tested the FGS on-sky. Figure~\ref{fig:FGS2023} shows the performance of the FGS during a 2023 on-sky pointing test from the ground. Atmospheric seeing limited the FGS stabilization to the 0.1\,$''$ (1-$\sigma$) level and moved the PSF on the FGC considerably from frame to frame. Nevertheless, the FGS gain parameters were tuned for the highbay environment, which allowed the FGS to stabilize at a level that is indicative of sub-arcsecond performance during flight. The gyroscope-limited performance at the FGS was at the 0.02--0.08\,$''$ (1-$\sigma$) level. Gyroscope-limited pointing occurs when the ACS preferentially weights the inertial measurements from the gyroscopes over optical centroiding with the FGS and star cameras, which quickly becomes seeing-limited on the ground. When the payload is at float altitudes and is not seeing-limited, the 1-$\sigma$ on-sky stability with star cameras is expected to become more aligned with the demonstrated gyroscope-limited pointing. A complete discussion of the EXCITE gondola pointing performance during the 2023 campaign is provided by~\citet{romualdez2024exoplanet}

The rest of the science instrument worked as expected. The cryogenic system maintained the required internal temperatures. We enabled the cryocooler vibration reduction system for the first time on the integrated instrument, verifying the functionality and performance of the active and passive vibration reduction modes over a range of operating conditions.~\cite{kirkconnell2024cryocooler} Science detector operation was verified, and we used the FSVC for the first time.

\subsection{\label{sec:2024}2024 Fort Sumner Flight}

\begin{figure}
    \centering
    \includesvg[width=\linewidth]{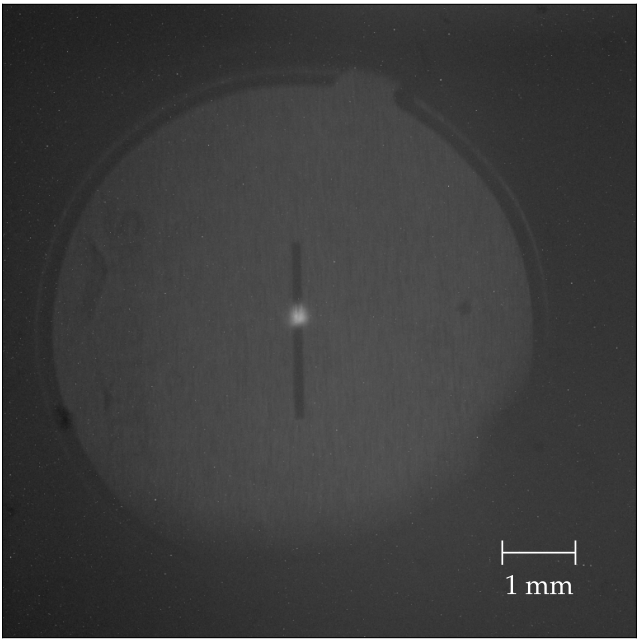}
    \caption{An image from the FSVC of the input field stop. The bright dot is a focused beam from a target star during on-sky pointing tests.}
    \label{fig:SVC_pic}
\end{figure}

EXCITE's 2024 I\&T campaign was largely uneventful, with just 17 days between payload arrival and flight readiness. Similar to the 2023 campaign, we demonstrated the seeing-limited ground pointing at the 1--2\,$''$ level using the star cameras. Likewise, with the FGS enabled, the stars on the FGC showed the same seeing-limited obfuscation down to 0.1\,$''$. The FSVC readout was streamlined and made for easier image capturing. This allowed us to more effectively aim the science beam onto the input field stop. Figure~\ref{fig:SVC_pic} shows an image from the FSVC with a focused beam directed onto the field stop of the spectrograph during a pre-flight on-sky pointing test.

\begin{figure}
    \centering
    \includegraphics[width=\linewidth]{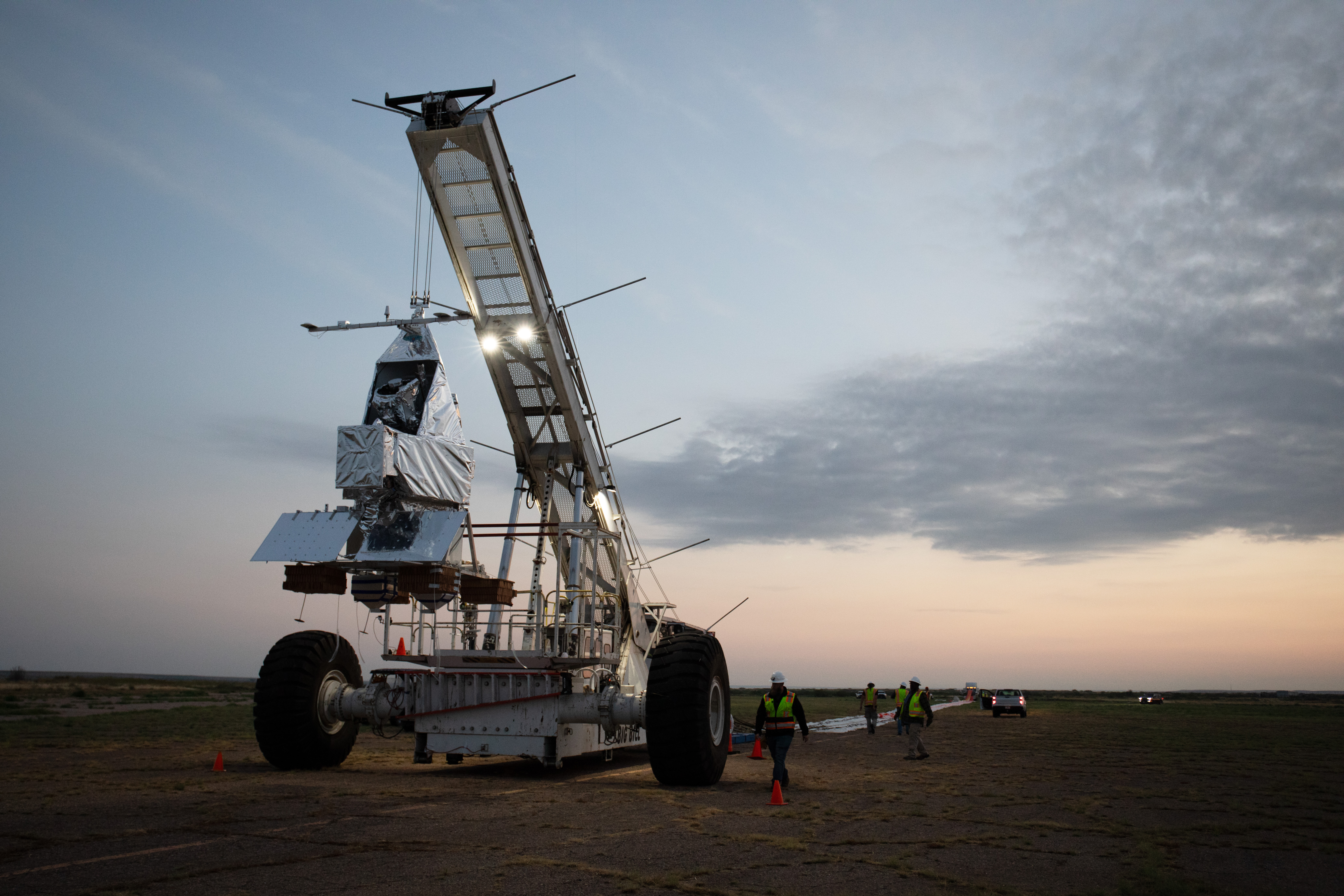}
    \caption{EXCITE on the launch vehicle at CSBF in Fort Sumner, New Mexico. The parachute can be seen being unrolled behind the payload and launch vehicle. Photo taken by Jeanette Kazmierczak/NASA.}
    \label{fig:exciteflightline}
\end{figure}

\begin{figure}
    \centering
    \includesvg[width=\linewidth]{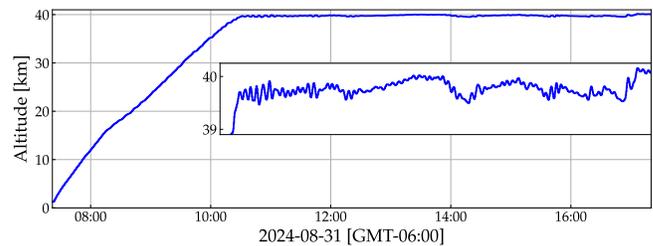}
    \caption{The GPS altitude during the EXCITE engineering flight. The zoomed-in region shows the altitude variations once EXCITE stabilized at float. Note that this GPS unit provided payload latitude, longitude, and altitude.}
    \label{fig:FTS2024_altitude}
\end{figure}

\begin{figure}
    \centering
    \includegraphics[width=\linewidth]{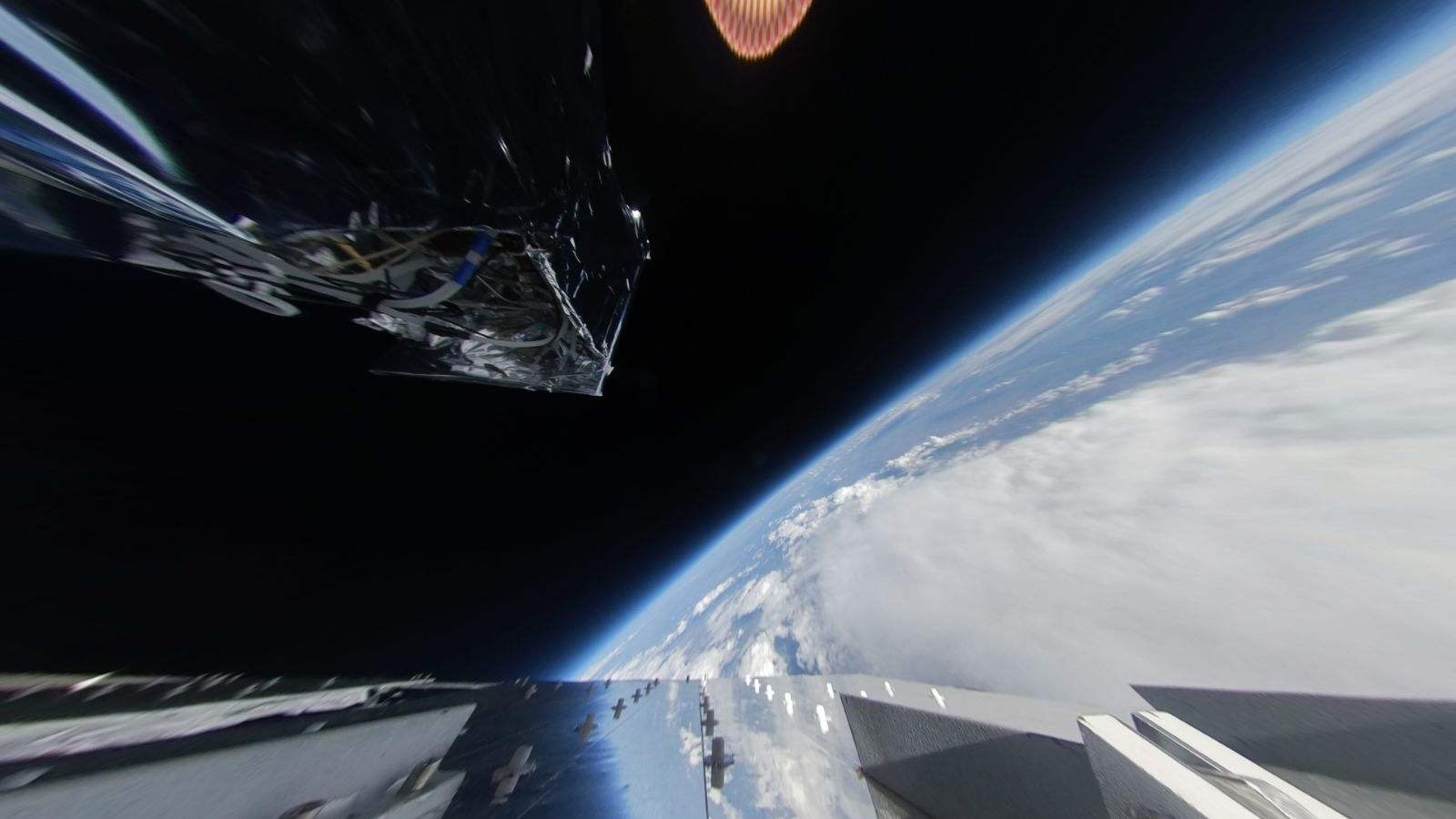}
    \caption{A photo of EXCITE at float altitudes captured by a 360$^{\circ}$ camera mounted to the front of the payload. The photo does not show the full 360$^{\circ}$ perspective. The radiator panels are seen at the bottom of the photo, and the parachute can be seen in orange at the top of the photo. An enclosure for the fluid loop pumps and tubing can be seen in the top left.}
    \label{fig:gopropic}
\end{figure}

EXCITE launched from Fort Sumner on August 31, 2024, at 07:22:00 local time (MDT/GMT-6:00). A photo of EXCITE on the flight line before launch is shown in Figure~\ref{fig:exciteflightline}. The flight ended at 17:35:31 MDT and lasted for 10~hours, 13~minutes, and 31~seconds. A plot of the GPS altitude data is shown in Figure~\ref{fig:FTS2024_altitude}. EXCITE reached a mean float altitude of 39.75\,km at around 10:30\,MDT. EXCITE had about 5\,hours at float altitudes, making a few rapid 0.25\,km oscillations during the first two hours at float before slowing down and oscillating 0.5\,km over the remaining 3\,hours. The payload did not see any significant drop in altitude during data acquisition, and even reached altitudes >\,40\,km about a half hour before termination. We mounted a 360$^{\circ}$ camera to the bow of the gondola, and Figure~\ref{fig:gopropic} shows a cropped image taken by this camera during the flight. EXCITE traveled $\sim$\,650\,km west of Fort Sumner and landed in the Fort Apache Reservation near Show Low, Arizona. The payload was recovered successfully with no significant damage incurred during flight, termination, or landing. Section~\ref{sec:pointingperformance} reports the performance of the gondola and ACS. Section~\ref{sec:cryoperformance} outlines the performance of the science instrument, including the cryogenic and detector systems.

\subsubsection{\label{sec:pointingperformance}Gondola \& ACS Flight Performance}

During flight, the gondola and ACS largely performed as expected. We were able to stabilize the gondola to <\,1\,$''$ during the flight. EXCITE demonstrated sub-arcsecond telescope stabilization over 15-minute periods when the payload was stabilizing with the 3-axes gimbal motors. This is an example of gyroscope-limited pointing defined in Section~\ref{sec:2023}. EXCITE stabilized at a gimbal position of 55.1$^{\circ}$ elevation, 0$^{\circ}$ roll, and $\sim$\,180$^{\circ}$ anti-Sun in azimuth and demonstrated 0.829\,$''$ (1-$\sigma$) stability measured by the gyroscopes. 

Unfortunately, the payload lost communication with the Trimble GPS compass immediately after launch, resulting in a loss of absolute azimuth information during the entire flight. This hindered our ability to track on pre-assigned sky targets, which we selected for their brightness and observational stability (both photometrically stable and viewable for for hours at a time within EXCITE's pointing constraints.) As a result, on-sky tracking of stars on the boresight and roll cameras lasted for only a few minutes before stars left our field, which was not enough time to enable the FGS and direct the beam into the spectrograph.

In addition, the angular travel of the elevation axis was limited to $\sim$\,45$^{\circ}$--57$^{\circ}$, instead of the nominal 22$^{\circ}$--57$^{\circ}$. This contributed to the relatively short amount of time where we were tracking on random sky targets. This would not have limited our ability to track on pre-assigned targets, which were selected to peak in the upper range of the telescope's elevation travel. We believe this problem was caused by a CTE mismatch between the frameless motor bearing housings (aluminum) and the bearings themselves (stainless steel). They were a press fit at room temperature; we believe that when the payload cooled during ascent through the troposphere, the housings contracted enough to plastically deform the bearings. We discuss re-engineering and testing of this system following the 2024 flight in Section~\ref{sec:refurb}. 

\subsubsection{\label{sec:cryoperformance}Science Instrument Flight Performance}
While the problems with the ACS described above prevented us from measuring a stellar spectrum, the science instrument otherwise performed nominally in flight. We tested many operating modes of the detector system, from windowed readout of spectral channels to full frame images at maximum bandwidth. We tested all mechanisms and actuators that are integral to the science instrument, including focus and tip/tilt of both the FGS and the telescope secondary mirror. If we had been able to lock on a target with the FGS, we are confident that the spectrometer would have performed the same in flight as on the ground.

\begin{figure}
    \centering
    \includesvg[width=\linewidth]{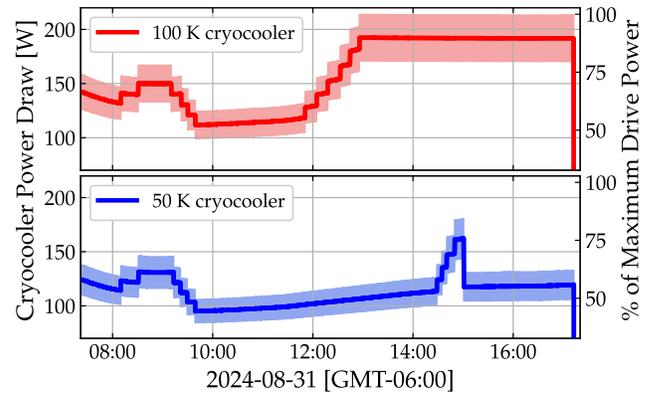}
    \caption{The estimated power sent to cryocoolers by WCS controllers during the flight. The shaded areas indicate 1-$\sigma$ uncertainty on the power due to the uncertainty in the supply voltage from the batteries on the payload during flight. The solid lines are the calculated power when using a supply voltage of 54\,V. The right-hand side $y$-axis shows the percent of the maximum allowable drive power to show their margins. The steps profiles correspond to commanded changes in the input power to the cryocoolers.}
    \label{fig:powerdraw}
\end{figure}

\begin{figure}
    \centering
    \includesvg[width=\linewidth]{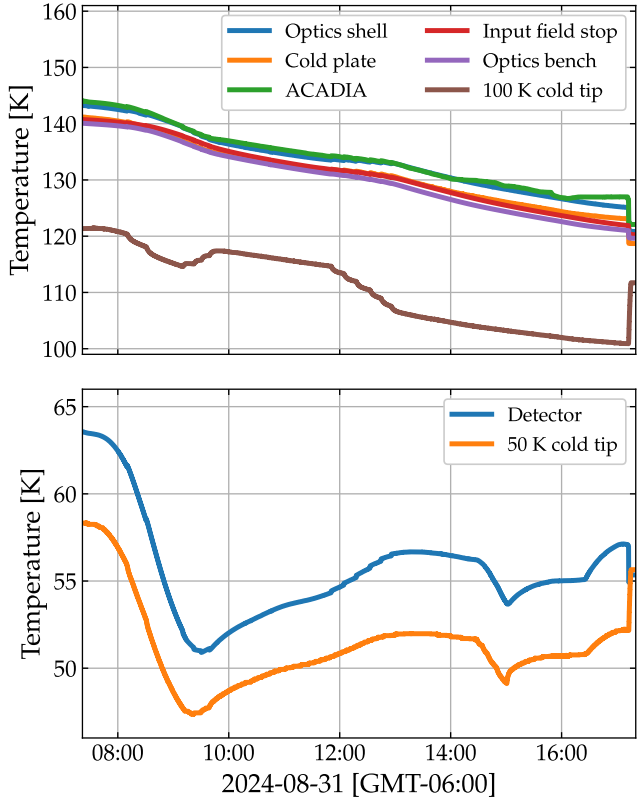}
    \caption{(Top) The temperatures of components at the 100\,K stage. (Bottom) The temperatures of the detector stage and the 50\,K cold tip.}
    \label{fig:internal_T}
\end{figure}

\begin{figure}
    \centering
    \includesvg[width=\linewidth]{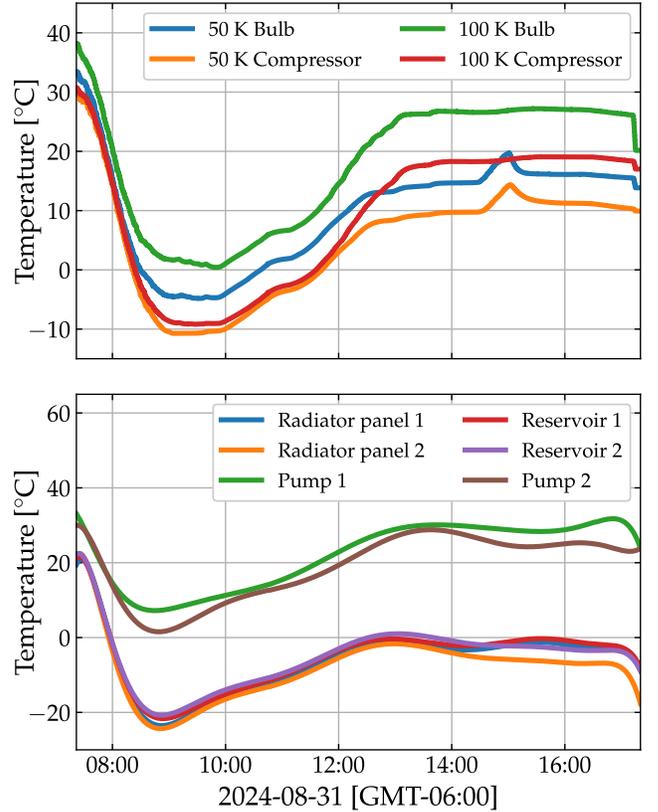}
    \caption{The temperatures of components external to the cryogenic receiver. (Top) The cryocooler skin temperatures, including the bulbs and compressors. (Bottom) The temperatures of the fluid loop components. Large drops in temperature occurred around 08:00 during ascent through the troposphere, where the ambient temperature drops to around -50\,$^{\circ}$C.}
    \label{fig:external_T}
\end{figure}

The cryogenic system was functional during the entire flight and met EXCITE's thermal requirements. We intentionally opted out of closed loop temperature control for the majority of flight, instead opting to explore the operational ranges of the cryocoolers to determine allowable margin and system behavior at different input powers. Figure~\ref{fig:powerdraw} shows the power drawn by the cryocoolers over the duration of the flight. The percent of maximum allowable drive power is also shown to highlight that the cryocoolers could have been driven to higher cooling powers. The temperatures during flight of the cold components inside the cryogenic receiver are shown in Figure~\ref{fig:internal_T}. Figure~\ref{fig:external_T} shows the temperatures of the warm components: the cryocooler skins, radiator panels, fluid pumps, and methanol reservoirs. The shapes of the temperature profiles in Figures~\ref{fig:internal_T} and~\ref{fig:external_T} can be correlated to the changes in input power to the cryocoolers; the cryogenic temperatures will decrease when the input power increases, and the cryocooler skin temperatures will increase when the input power increases.

To avoid overheating the cryocooler skins while on the launch pad, the cryocoolers were not driven to their nominal operating power, and the temperatures of the optics shell and the ACADIA began the flight at 143\,K. The 100\,K stage continued to cool during the flight, and once we determined that the skin temperatures were stable and adjusted to the ambient temperature at float, we increased the input power to the 100\,K cryocooler to 180\,W. The ACADIA was PID-controlled at 127\,K during the last stage of the flight, and achieved a 1-$\sigma$ temperature stability of 0.01\,K for 30~minutes. By the end of the flight, the 100\,K cold tip had cooled to 100.9\,K, drawing $\sim$\,190\,W. The optics shell temperature cooled to as low as 125.0\,K at less than 90\% of maximum operating power.

The detector stayed below 57.5\,K during the entire flight once it reached float. Its cryocooler drew on average about 110\,W, about 50--60\% of its maximum drive power. At the temperatures measured during flight, the dark current on the H2RG is negligible for EXCITE.

The cryocooler control electronics' vibration-reduction systems functioned nominally during the flight. Exported vibration levels measured by the accelerometers in flight were comparable to those we measured on the ground, and we measured no negative effects to the pointing system due to vibrations from the cryocooler compressors.

The heat dissipation scheme for the cryocoolers also worked well. The cryocooler skin temperatures remained slightly above ambient for the entire flight, well within their safe operating range. The bulb of the 100\,K cryocooler, the warmest component between the two cryocoolers, reached 27\,$^{\circ}$C at its largest input power during the flight. The radiator panels stayed below 0\,$^{\circ}$C. The radiator panels showed no signs of azimuthal dependency as the payload faced towards the Sun during points of the flight. Even as we increased the input power to the cryocoolers, the radiator panels maintained stable temperatures. The fluid pumps warmed to between $\sim$\,25--28\,$^{\circ}$C, well within their operational temperature limits, and operated normally for the entire flight.

\subsection{\label{sec:refurb}Refurbishments}

To prepare for a long-duration science flight, we have made several refurbishments and upgrades to EXCITE. Motivated by the loss of the GPS compass, for future flights we plan to utilize pinhole sun sensors to provide redundant azimuth determination. If the GPS compass were to fail again, we can use the sun sensors to orient the azimuth of the payload to <\,0.5$^{\circ}$ accuracy, and feed that information back to the magnetometer and assist the star cameras in solving the payload's absolute attitude. The sun sensor design is based on the ones flown on the BLASTPol~\cite{fissel2010balloon} and EBEX~\cite{aboobaker2018ebex} missions. Following the design by Korotkov \textit{et al.},\cite{korotkov2013pinhole} the sun sensor operates as a pinhole camera, projecting the Sun’s image through a 200\,\textmu m diameter circular aperture onto a 9\,mm\,$\times$\,9\,mm position-sensitive detector (PSD). By detecting the image position, the sensor determines its orientation relative to the Sun. With a 40$^{\circ}$ field of view and fixed mounting on the gondola, it provides azimuth and elevation angles of the Sun relative to the payload. EXCITE will operate three sun sensors. These units have been tested at the component level and will be integrated with the payload before the science flight. We calibrated the sun sensors by tracking the position of the Sun at an altitude of about 25\,m, and they have demonstrated an accuracy of 0.24$^{\circ}$ rms in measuring the Sun's position in both elevation and azimuth. This results in an accuracy of 0.35$^{\circ}$ rms in measuring the absolute position of the Sun.

To remedy the CTE mismatch in the frameless motors, we remachined the bearing housings from steel. We then tested the elevations drives under representative loading over a wide range of operating conditions. To achieve the correct loading, we fabricated a ``mock inner frame,'' composed of the frameless motors, the coarse elevation stepper motors, and 225\,kg of steel and aluminum sandwiched between the two sets of motors in order to simulate the mass of the science instrument. We then tested the assembly in a TVAC chamber at CSBF in Palestine, Texas. In this test, the mock inner frame was actuated through the full elevation range at pressures spanning 4--800\,mbar and at ambient temperatures between -60\,$^{\circ}$C and 40\,$^{\circ}$C, including a cold soak at -60\,$^{\circ}$C for 1~hour. The elevation motors operated as expected at all pressures and temperatures during a simulated ascent through the troposphere.

Additionally, we cleaned the primary and secondary mirrors to remove any contaminants that accumulated on their surfaces during flight and after landing. After subsequent re-alignment of the telescope optics, we are continuing to perform full system tests, such as on-sky pointing and thermal cycling the cryogenic system. We constructed an artificial source in order to enable the ACS in the case of poor outdoor seeing, and we anticipate it will aid us in aligning the telescope beam onto the detector focal plane. To reduce electromagnetic interference (EMI), we rebuilt our cryocooler power supplies to include EMI filters. We also integrated the solar panel arrays, which EXCITE will fly during an LDB flight. 

\section{\label{sec:summary}Summary and Outlook}

Characterizing exoplanet atmospheres is a rapidly growing field. The methods and simulation efforts to quantify chemical composition, heat dynamics, and energy recirculation on highly-irradiated hot Jupiters are well-known and practiced. Despite these efforts, the amount of observational data to constrain this large parameter space is lacking. White light phase curves of hot Jupiters are abundant, but do not probe the three-dimensional structure of the exoplanet atmosphere. Similarly, measurements that are taken during only transit and only through secondary eclipse are limited since they do not spatially resolve the exoplanet atmosphere, can suffer from degeneracies when disentangling chemical abundances from thermal emission, measure only the limb of the atmosphere (for transits), and provide only hemispherical averages (for secondary eclipses). Full-orbit spectroscopic phase curves solve these problems, but they are challenging and resource-intensive observations to make. Only a handful of spectroscopic phase curves exist, and all require shared space-based observatories. EXCITE will be the first instrument dedicated to making these observations, with the capability to double to the number of spectroscopic phase curves that exist. In the process, EXCITE will be the first instrument verify the observational and systematic detrending techniques required to obtain phase-resolved spectra from a balloon platform.

We report the mechanical, optical, and thermal design for the EXCITE payload, as well as results from radiometric simulations, jitter simulations, spectral image generation, and data reduction tools. We show that EXCITE is expected to be photon-noise limited by the target across the majority of the passband, and we expect jitter noise to be subdominant to Earth's atmosphere and thermal emission. We understand the systematic environment from a correlated and uncorrelated noise perspective, and we show that EXCITE can meet its pointing and thermal requirements with margin from a balloon environment. EXCITE flew an engineering flight from Fort Sumner, New Mexico in August 2024. The gondola and ACS delivered sub-arcsecond stability. We could not track stars for hours-long due to a loss in GPS compass, but the star cameras were able to track stars for $\sim$\,30~minute durations. The cryogenic system met all thermal requirements with significant margin on the input power to the cryocoolers. The cryostat cooled the detector to 55\,K and the optics shell to 125\,K, and the heat rejection scheme worked well. The detector system and all onboard electronics performed nominally. We have since made refurbishments and upgrades to the payload to fix the problems encountered during flight. 
 
We are currently preparing for an LDB flight in the 2026-2027 Austral Summer. EXCITE will deliver high-quality spectroscopic phase curves and significantly increase the available spectroscopic observational data on hot Jupiters. EXCITE aims to set the standard for phase curve observing and exoplanet atmosphere characterization from a balloon platform.

\begin{acknowledgments}
This work is supported by the National Aeronautics and Space Administration under Award No. 18-APRA18-0075 selected under NASA Research Announcement Solicitation No. NNH18ZDA001N, Research Opportunities in Space and Earth Sciences – 2018 (ROSES-2018).
\end{acknowledgments}

\bibliography{excite}
\bibliographystyle{aipnum4-1.bst}

\end{document}